\documentclass{article}
\usepackage{amssymb}
\usepackage{amsmath}

\input{tcilatex}
\begin{document}

\begin{center}

{\LARGE Properties of the zeros of generalized hypergeometric polynomials}

\bigskip

$^{\ast }$\textbf{Oksana Bihun}$^{1}$ and $^{+\lozenge }$\textbf{Francesco
Calogero}$^{2}\bigskip $

$^{\ast }$Department of Mathematics, Concordia College\\
 901 8th Str. S,  Moorhead, MN 56562, USA, +1-218-299-4396
\smallskip

$^{+}$Physics Department, University of Rome \textquotedblleft La Sapienza"\\
 p. Aldo Moro, I-00185 ROMA, Italy, +39-06-4991-4372
\smallskip

$^{\lozenge }$Istituto Nazionale di Fisica Nucleare, Sezione di Roma
\smallskip

$^{1}$Corresponding author, obihun@cord.edu

$^{2}$francesco.calogero@roma1.infn.it, francesco.calogero@uniroma1.it

\bigskip

\textit{Abstract}
\end{center}

We define the \textit{generalized hypergeometric polynomial} of degree $N$
as follows:%
\begin{eqnarray*}
&&P_{N}\left( \alpha _{1},...,\alpha _{p};\beta _{1},...,\beta _{q};z\right)
=\sum_{m=0}^{N}\left[ \frac{\left( -N\right) _{m}\left( \alpha _{1}\right)
_{m}\cdot \cdot \cdot \left( \alpha _{p}\right) _{m}~z^{N-m}}{m!~\left(
\beta _{1}\right) _{m}\cdot \cdot \cdot \left( \beta _{q}\right) _{m}}\right]
\\
&=&z^{N}~_{p+1}F_{q}\left( -N,\alpha _{1,}...,\alpha _{p};\beta
_{1},...,\beta _{q};1/z\right) ~.
\end{eqnarray*}%
Here $N$ is an arbitrary \textit{positive} integer, $p$ and $q$ are
arbitrary \textit{nonnegative} integers, the $p+q$ parameters $\alpha _{j}$
and $\beta _{k}$ are arbitrary ("generic", possibly complex) numbers, $%
\left( \alpha \right) _{m}$ is the Pochhammer symbol and $_{p+1}F_{q}\left(
\alpha _{0},\alpha _{1},...\alpha _{p};\beta _{1},...,\beta _{q};z\right) $
is the generalized hypergeometric function. In this paper we obtain a set of 
$N$ \textit{nonlinear algebraic equations} satisfied by the $N$ zeros $\zeta
_{n}$ of this polynomial. We moreover manufacture an $N\times N$ matrix $%
\underline{L}$ in terms of the $1+p+q$ parameters $N$, $\alpha _{j}$, $\beta
_{k}$ characterizing this polynomial, and of its $N$ zeros $\zeta _{n}$, and
we show that it features the $N$ eigenvalues $\lambda
_{m}=m~\tprod\nolimits_{k=1}^{q}\left( -\beta _{k}+1-m\right)
~,~~~m=1,...,N~ $. These $N$ eigenvalues depend only on the $q$ parameters $%
\beta _{k}$, implying that the $N\times N$ matrix $\underline{L}$ is \textit{%
isospectral} for variations of the $p$ parameters $\alpha _{j}$; and they
clearly are \textit{integer} (or \textit{rational}) numbers if the $q$
parameters $\beta _{k}$ are themselves \textit{integer }(or \textit{rational}%
) numbers: a nontrivial \textit{Diophantine} property.



\bigskip

\section{Introduction}

The investigation of the properties of the zeros of polynomials has a long
history going back for several centuries, yet new approaches and findings
have also emerged in relatively recent times. A class of such findings
extended the results pioneered by G. Sz\"{e}go (see in particular Section
6.7 of \cite{S1939}), by identifying additional sets of \textit{nonlinear
algebraic relations} satisfied by the zeros of the classical polynomials
and, more generally, of polynomials belonging to the Askey scheme, as well
as $N\times N$ matrices, constructed with the $N$ zeros of these polynomials
(of degree $N$), whose eigenvalues could be explicitly identified and in
many cases feature \textit{Diophantine} properties: see for instance \cite%
{ABCOP1979} \cite{BCD1} \cite{BCD2} \cite{BCD3} \cite{BCD4} \cite{CI2013} \cite{IR2013}. The present paper
reports analogous results for \textit{generalized hypergeometric polynomials}%
. These findings are displayed in the following Section 2, and proven in the
subsequent Section 3; certain polynomial identities which are essential for
obtaining and reporting these results---and are themselves remarkable---are
proven and displayed in Appendix A. A terse Section 4 ("Outlook") outlines
possible future developments.

\bigskip

\section{Results}

The generalized hypergeometric function $_{p+1}F_{q}\left( \alpha
_{0},\alpha _{1},...,\alpha _{p};\beta _{1},...,\beta _{q};z\right) $ is
defined as follows (see for instance \cite{HTF1}): 
\begin{subequations}
\begin{equation}
_{p+1}F_{q}\left( \alpha _{0},\alpha _{1},...,\alpha _{p};\beta
_{1},...,\beta _{q};z\right) =\sum_{j=0}^{\infty }\left[ \frac{\left( \alpha
_{0}\right) _{j}~\left( \alpha _{1}\right) _{j}\cdot \cdot \cdot \left(
\alpha _{p}\right) _{j}~z^{j}}{j!~\left( \beta _{1}\right) _{j}\cdot \cdot
\cdot \left( \beta _{q}\right) _{j}}\right] ~.  \label{GenHyp}
\end{equation}%
Above and throughout, the Pochhammer symbol $\left( \alpha \right) _{j}$ is
defined as follows:%
\begin{equation}
\left( \alpha \right) _{0}=1~;~~~\left( \alpha \right) _{j}=\alpha ~\left(
\alpha +1\right) \cdot \cdot \cdot \left( \alpha +j-1\right) =\frac{\Gamma
\left( \alpha +j\right) }{\Gamma \left( \alpha \right) }~~~\text{for}%
~~~j=1,2,3,...~.  \label{Poch}
\end{equation}%
Clearly if one of the $p+1$ parameters $\alpha _{j}$ is a \textit{negative}
integer, say $\alpha _{0}=-N$, and all the other $p+q$ parameters $\alpha
_{j}$ and $\beta _{k}$ have \textit{generic} (possibly complex) values, the
series in the right-hand side of the definition (\ref{GenHyp}) of the
generalized hypergeometric function terminates at $j=N$ (since $\left(
-N\right) _{j}=0$ for $j=N+1,~N+2,...$) . Hereafter we call \textit{%
generalized hypergeometric polynomial} the resulting polynomial (of degree $%
N $ in $z$, and conveniently defined as follows, so that it is \textit{monic}%
): 
\end{subequations}
\begin{subequations}
\label{Poly}
\begin{equation}
P_{N}\left( \alpha _{1},...,\alpha _{p};\beta _{1},...,\beta _{q};z\right)
=\sum_{m=0}^{N}\left[ \frac{\left( -N\right) _{m}\left( \alpha _{1}\right)
_{m}\cdot \cdot \cdot \left( \alpha _{p}\right) _{m}~z^{N-m}}{m!~\left(
\beta _{1}\right) _{m}\cdot \cdot \cdot \left( \beta _{q}\right) _{m}}\right]
~;  \label{PN}
\end{equation}%
and we denote its $N$ zeros as $\zeta _{n}$, 
\begin{equation}
P_{N}\left( \alpha _{1},...,\alpha _{p};\beta _{1},...,\beta _{q};\zeta
_{n}\right) =0~,~~~n=1,...,N~.  \label{zeros}
\end{equation}%
Hence the values of the $N$ numbers $\zeta _{n}$ depend on the $1+p+q$
parameters $N,$ $\alpha _{j}$, $\beta _{k}$.

\textit{Notation 2.1}. Above and hereafter $N$ is an (arbitrarily assigned) 
\textit{positive} integer, $p\ $and $q$ are two (arbitrarily assigned) 
\textit{nonnegative} integers, and indices such as $n,$ $m,$ $\ell $ (but
not necessarily $j,$ $k$) run over the $N$ integers from $1$ to $N$ (unless
otherwise indicated). The $N$ zeros $\zeta _{n}$ are of course defined up to
permutations. In the following we always assume the \textit{same} assignment
to be made for the correlation of the values of the $N$ zeros of a
polynomial with the values of the index $n$ labeling them. And below
underlined lower-case letters denote $N$-vectors (hence, for instance, $%
\underline{\zeta }\equiv \left( \zeta _{1},...,\zeta _{N}\right) $); and
underlined upper-case letters denote $N\times N$ matrices (hence for
instance the matrix $\underline{L}$ has the $N^{2}$ elements $L_{nm}$).
Finally: we always adopt the standard convention according to which a sum
containing no terms vanishes, and a product containing no terms equals
unity: for instance, $\sum_{j,k=1;j\neq k}^{1}=0,$ $\prod\nolimits_{j,k=1;j%
\neq k}^{1}=1.$ $\square $

The first result of this paper consists of the following

\textit{Proposition 2.1}. The (set of) $N$ zeros $\zeta _{n}$ of the \textit{%
generalized hypergeometric polynomial} $P_{N}\left( \alpha _{1},...,\alpha
_{p};\beta _{1},...,\beta _{q};z\right) $, see (\ref{PN}) and (\ref{zeros}),
satisfy the following system of $N$ algebraic equations: 
\end{subequations}
\begin{equation}
\sum_{k=1}^{q+1}\left[ b_{k}~\ f_{n}^{\left( k\right) }\left( \underline{%
\zeta }\right) \right] -\sum_{j=0}^{p}\left[ a_{j}~g_{n}^{\left( j\right)
}\left( \underline{\zeta }\right) \right] =0~,~~~n=1,...,N~.  \label{Eqzitan}
\end{equation}%
Here the $q+1$ coefficients $b_{k}$, respectively the $p+1$ coefficients $%
a_{j}$, are defined in terms of the $q$ parameters $\beta _{k}$ respectively
the $p$ parameters $\alpha _{j}$ so that 
\begin{subequations}
\label{Paramb}
\begin{equation}
x~\tprod\limits_{k=1}^{q}\left( \beta _{k}-1-x\right)
=\sum_{k=1}^{q+1}\left( b_{k}~x^{k}\right) ~,
\end{equation}%
hence 
\begin{equation}
b_{1}=\tprod\limits_{k=1}^{q}\left( \beta _{k}-1~\right) ~,
\end{equation}%
\begin{equation}
b_{2}=-\sum_{j=1}^{q}\left[ \tprod\limits_{k=1,~k\neq j}^{q}\left( \beta
_{k}-1\right) \right] ~,
\end{equation}%
\begin{equation}
b_{3}=\frac{1}{2}\sum_{\ell ,j=1;\ell \neq j}^{q}\left[ \tprod%
\limits_{k=1,~k\neq \ell ,j}^{q}\left( \beta _{k}-1\right) \right] ~,
\end{equation}%
and so on up to%
\begin{equation}
b_{q+1}=\left( -1\right) ^{q}~;
\end{equation}%
respectively 
\end{subequations}
\begin{subequations}
\label{Parama}
\begin{equation}
\tprod\limits_{j=1}^{p}\left( \alpha _{j}-x\right)
=\sum_{j=0}^{p}a_{j}~x^{j}~,
\end{equation}%
hence%
\begin{equation}
a_{0}=\tprod\limits_{j=1}^{p}\left( \alpha _{j}\right) ~,
\end{equation}%
\begin{equation}
a_{1}=-\sum_{k=1}^{p}\left[ \tprod\limits_{j=1;~j\neq k}^{p}\left( \alpha
_{j}\right) \right] ~,
\end{equation}%
\begin{equation}
a_{2}=\frac{1}{2}\sum_{\ell ,k=1;\ell \neq k}^{p}\left[ \tprod%
\limits_{j=1;~j\neq \ell ,k}^{p}\left( \alpha _{j}\right) \right] ~,
\end{equation}%
and so on up to%
\begin{equation}
a_{p}=\left( -1\right) ^{p}~.
\end{equation}

As for the functions $f_{n}^{\left( j\right) }\left( \underline{\zeta }%
\right) $ of the $N$ zeros $\zeta _{m}$, they are defined recursively as
follows: 
\end{subequations}
\begin{subequations}
\label{RecDeffnj}
\begin{equation}
f_{n}^{\left( j+1\right) }\left( \underline{\zeta }\right) =-f_{n}^{\left(
j\right) }\left( \underline{\zeta }\right) +\sum_{\ell =1;~\ell \neq n}^{N}%
\left[ \frac{\zeta _{n}~f_{\ell }^{\left( j\right) }\left( \underline{\zeta }%
\right) +\zeta _{\ell }~f_{n}^{\left( j\right) }\left( \underline{\zeta }%
\right) }{\zeta _{n}-\zeta _{\ell }}\right] ~,~~~j=1,2,3,...~,
\end{equation}%
with%
\begin{equation}
f_{n}^{\left( 1\right) }\left( \underline{\zeta }\right) =\zeta _{n}~,
\end{equation}%
implying the expressions of $f_{n}^{\left( j\right) }\left( \underline{\zeta 
}\right) $ with $j=1,2,3,...$ reported in the Appendix, see (\ref{ff}).

And the functions $g_{n}^{\left( j\right) }\left( \underline{\zeta }\right) $
of the $N$ zeros $\zeta _{n}$ are defined as follows: 
\end{subequations}
\begin{subequations}
\label{gnj}
\begin{equation}
g_{n}^{\left( 0\right) }\left( \underline{\zeta }\right) =1~,
\end{equation}%
\begin{equation}
g_{n}^{\left( j\right) }\left( \underline{\zeta }\right) =\sum_{\ell
=1;~\ell \neq n}^{N}\left[ \frac{f_{n}^{\left( j\right) }\left( \underline{%
\zeta }\right) +f_{\ell }^{\left( j\right) }\left( \underline{\zeta }\right) 
}{\zeta _{n}-\zeta _{\ell }}\right] ~,~~~j=1,2,...~,
\end{equation}%
implying%
\begin{equation}
g_{n}^{\left( 1\right) }\left( \underline{\zeta }\right) =\sum_{\ell
=1;~\ell \neq n}^{N}\left( \frac{\zeta _{n}+\zeta _{\ell }}{\zeta _{n}-\zeta
_{\ell }}\right) ~,
\end{equation}%
and the expressions of $g_{n}^{\left( j\right) }\left( \underline{\zeta }%
\right) $ with $j=1,2,3,...$ reported in the Appendix, see (\ref{ggg}).~$%
\square $

\textit{Remark 2.1}. The functions $f_{n}^{\left( j\right) }\left( 
\underline{\zeta }\right) $ and $g_{n}^{\left( j\right) }\left( \underline{%
\zeta }\right) $ are \textit{universal}: they do not depend on the
generalized hypergeometric polynomial under consideration. But of course
their arguments do, being the $N$ zeros $\zeta _{n}$ of the polynomial $%
P_{N}\left( \alpha _{1},...,\alpha _{p};\beta _{1},...,\beta _{q};\zeta
\right) ,$ see (\ref{Poly}). $\square $

The second (and main) result of this paper consist of the following

\textit{Proposition 2.2.} Let the (unordered) set of $N$ numbers $\zeta _{n}$
denote the $N$ zeros of the \textit{generalized hypergeometric polynomial} $%
P_{N}\left( \alpha _{1},...,\alpha _{p};\beta _{1},...,\beta _{q};z\right) $%
, see (\ref{PN}) and (\ref{zeros}); and let the $N\times N$ matrix $%
\underline{L}\left( \underline{\zeta }\right) $ be defined componentwise as
follows, in terms of these $N$ zeros and the $1+p+q$ parameters $N,$ $\alpha
_{j}$, $\beta _{k}$ characterizing the \textit{generalized hypergeometric
polynomial} $P_{N}\left( \alpha _{1},...,\alpha _{p};\beta _{1},...,\beta
_{q};z\right) $: 
\end{subequations}
\begin{equation}
L_{nm}\left( \underline{\zeta }\right) =\sum_{k=1}^{q+1}\left[
b_{k}~f_{n,m}^{\left( k\right) }\left( \underline{\zeta }\right) \right]
-\sum_{j=1}^{p}\left[ a_{j}~g_{n,m}^{\left( j\right) }\left( \underline{%
\zeta }\right) \right] ~,  \label{Mnmtilde}
\end{equation}%
where of course the coefficients $b_{k}$ and $a_{j}$ are defined as above,
see (\ref{Parama}) and (\ref{Paramb}), while $f_{n,m}^{\left( k\right)
}\left( \underline{\zeta }\right) $ respectively $g_{n,m}^{\left( j\right)
}\left( \underline{\zeta }\right) $ are defined, in terms of the quantities $%
f_{n}^{\left( k\right) }\left( \underline{\zeta }\right) $ respectively $%
g_{n}^{\left( j\right) }\left( \underline{\zeta }\right) $ (see (\ref%
{RecDeffnj}) respectively (\ref{gnj})), as follows: 
\begin{equation}
f_{n,m}^{\left( k\right) }\left( \underline{\zeta }\right) =\left. \frac{%
\partial ~f_{n}^{\left( k\right) }\left( \underline{z}\right) }{\partial
~z_{m}}\right\vert _{\underline{z}=\underline{\zeta }}~,~~~g_{n,m}^{\left(
j\right) }\left( \underline{\zeta }\right) =\left. \frac{\partial
~g_{n}^{\left( j\right) }\left( \underline{z}\right) }{\partial ~z_{m}}%
\right\vert _{\underline{z}=\underline{\zeta }}~.  \label{fgjnm}
\end{equation}%
Expressions of $f_{n,m}^{\left( j\right) }\left( \underline{\zeta }\right) $
and $g_{n,m}^{\left( j\right) }\left( \underline{\zeta }\right) $ with $%
j=1,2,3,...$ are reported in the Appendix, see (\ref{fnm}) and (\ref{gnm}).

Then the $N$ eigenvalues $\lambda _{m}$ of the $N\times N$ matrix $%
\underline{L}\left( \underline{\zeta }\right) $, 
\begin{subequations}
\begin{equation}
\underline{L}\left( \underline{\zeta }\right) ~\underline{v}^{\left(
m\right) }\left( \underline{\zeta }\right) =\lambda _{m}~\underline{v}%
^{\left( m\right) }\left( \underline{\zeta }\right) ~,~~~m=1,...,N
\label{EigenEqMtilde}
\end{equation}%
---hence the $N$ roots $\lambda _{m}$ of the following polynomial equation
(of degree $N$ in $\lambda $):%
\begin{equation}
\det \left[ \underline{L}\left( \underline{\zeta }\right) -\lambda \right] =0
\end{equation}%
---are given by the formula 
\begin{equation}
\lambda _{m}\left( \beta _{1},...,\beta _{q}\right)
=m~\tprod\limits_{k=1}^{q}\left( \beta _{k}-1+m\right)
~,~~~m=1,...,N~.~\square  \label{mumtilde}
\end{equation}%
\qquad

\textit{Remark 2.2}. The functions $f_{n,m}^{\left( j\right) }\left( 
\underline{\zeta }\right) $ and $g_{n,m}^{\left( j\right) }\left( \underline{%
\zeta }\right) $ are \textit{universal}: they do not depend on the
generalized hypergeometric polynomial under consideration (see \textit{%
Remark 2.1}). But of course their arguments do, being the $N$ zeros $\zeta
_{n}$ of the polynomial $P_{N}\left( \alpha _{1},...,\alpha _{p};\beta
_{1},...,\beta _{q};\zeta \right) ,$ see (\ref{Poly}). $\square $

\textit{Remark 2.3}. The $N$ eigenvalues $\lambda _{m}$ of the $N\times N$
matrix $\underline{L}$ (see (\ref{Mnmtilde})) depend only on the $q$
parameters $\beta _{k}$ (see (\ref{mumtilde})), while the matrix $\underline{%
L}$ depends itself on the $q+p$ parameters $\beta _{k}$ and $\alpha _{j}$%
---via the dependence of the parameters $b_{k}$ respectively $a_{j}$ on $%
\beta _{k}$ respectively $\alpha _{j}$ (see (\ref{Paramb}) respectively (\ref%
{Parama})) and the dependence of the $N$ zeros $\zeta _{n}$ on the
parameters $\beta _{k}$ and $\alpha _{j}$, see (\ref{zeros}) or,
equivalently, on $b_{k}$ and $a_{j}$, see (\ref{Eqzitan}). Hence the $%
N\times N$ matrix $\underline{L}$ is \textit{isospectral} for variations of
the $p$ parameters $\alpha _{j}$. And note moreover that the $N$ eigenvalues 
$\lambda _{m}$ are \textit{integer} (or \textit{rational}) numbers if the $q$
parameters $\beta _{k}$ are themselves \textit{integer }(or \textit{rational}%
) numbers: a nontrivial \textit{Diophantine} property of the $N\times N$
matrix $\underline{L}$.~$\square $

\textit{Remark 2.4}. All the above results are of course true as written
only provided the $N$ zeros $\zeta _{n}$ are all different among themselves;
but they clearly remain valid by taking appropriate limits whenever this
restriction does not hold. $\square $

\textit{Remark 2.5}. Immediate generalizations---whose explicit formulations
can be left to the interested reader--- of \textit{Propositions 2.1} and 
\textit{2.2 }obtain from these two propositions via the special assignment $%
\alpha _{\hat{q}+j}=\beta _{\hat{p}+j}$ for $j=1,...r$ with $r$ an arbitrary 
\textit{nonnegative integer} such that both $\hat{q}=q-r$ and $\hat{p}=p-r$
are \textit{positive integers}. These propositions refer then to the $N$
zeros of the generalized hypergeometric polynomial $P_{N}\left( \alpha
_{1},...,\alpha _{\hat{p}};\beta _{1},...,\beta _{\hat{q}};z\right) $%
---which depend only on the $1+\hat{p}+\hat{q}=1+p+q-2r$ parameters $N,$ $%
\alpha _{j}$ with $j=1,...,\hat{p}=p-r$ and $b_{k}$ with $k=1,...,\hat{q}%
=q-r $, but feature quantities $b_{k}$ and $a_{j}$ (see (\ref{Eqzitan}) and (%
\ref{Mnmtilde})) that depend on the $1+p+q$ parameters $N,$ $\alpha _{j}$
with $j=1,...,p$ and $b_{k}$ with $k=1,...,q$. $\square $

The two \textit{Propositions 2.1} and 2.\textit{2} are proven in the
following Section 3; some comments and prospects of future developments are
outlined in the last Section 4.

Let us end this Section 2 by displaying explicitly the above results for
small values of the integers $p$, $q$ and (of course) $r$ (see \textit{%
Remark 2.5}).

\subsection{The case $p=q=1,$ $r=0$}

For $p=q=1,$ $r=0$ (for the definition of $r$ see \textit{Remark 2.5})
implying (see (\ref{Parama}) and (\ref{Paramb})) 
\end{subequations}
\begin{equation}
a_{0}=\alpha _{1},~a_{1}=-1,~~~b_{1}=\beta _{1}-1,~b_{2}=-1~,
\end{equation}%
let the $N$ numbers $\zeta _{n}$ be the $N$ zeros of the hypergeometric
polynomial%
\begin{equation}
P_{N}\left( \alpha _{1};\beta _{1};z\right) =\sum_{m=0}^{N}\left[ \frac{%
\left( -N\right) _{m}\left( \alpha _{1}\right) _{m}~z^{N-m}}{m!~\left( \beta
_{1}\right) _{m}}\right] ~.  \label{PNp=q=1}
\end{equation}%
Then \textit{Proposition 2.1} implies that these $N$ zeros $\zeta _{n}$
satisfy the following system of $N$ nonlinear algebraic equations%
\begin{equation}
N-1-\alpha _{1}+\beta _{1}~\zeta _{n}-2~\left( \zeta _{n}-1\right) ~\sigma
_{n}^{\left( 1,1\right) }\left( \underline{\zeta }\right) =0~,~~~n=1,...,N~.
\label{12c}
\end{equation}

\textit{Notation 2.2}. Above and hereafter the notation $\sigma _{n}^{\left(
r,\rho \right) }\left( \underline{\zeta }\right) $ is defined by (\ref%
{sigmanpq}). $\square $

As for \textit{Proposition 2.2,} it implies in this case that the following $%
N\times N$ matrix\textbf{\ }$\underline{L}\left( \underline{\zeta }\right) $%
, defined componentwise as follows (see (\ref{Mnmtilde}) with (\ref{fnm}), (%
\ref{gnm}) and (\ref{sigmanpq})), 
\begin{subequations}
\begin{eqnarray}
L_{nm}\left( \underline{\zeta }\right) &=&\delta _{nm}~\left\{ \beta
_{1}+2~\sum_{\ell =1;\ell \neq n}^{N}\left[ \frac{\zeta _{\ell }~\left(
\zeta _{\ell }-1\right) }{\left( \zeta _{n}-\zeta _{\ell }\right) ^{2}}%
\right] \right\}  \notag \\
&&-2~\left( 1-\delta _{nm}\right) ~\frac{\zeta _{n}~\left( \zeta
_{n}-1\right) }{\left( \zeta _{n}-\zeta _{m}\right) ^{2}}~,  \label{12d}
\end{eqnarray}%
features the $N$ eigenvalues 
\begin{equation}
\lambda _{m}=m~\left( \beta _{1}-1+m\right) ~,~~~m=1,...,N~.  \label{12e}
\end{equation}

\subsection{The case $p=2$, $q=1,$\ $r=0$}

For $p=2$, $q=1,$\ $r=0$ (for the definition of $r$ see \textit{Remark 2.5}%
), implying (see (\ref{Parama}) and (\ref{Paramb})) 
\end{subequations}
\begin{equation}
a_{0}=\alpha _{1}~\alpha _{2},~a_{1}=-\left( \alpha _{1}+\alpha _{2}\right)
,~a_{2}=1~,~~~b_{1}=\beta _{1}-1,~b_{2}=-1,\ 
\end{equation}%
let the $N$ numbers $\zeta _{n}$ be the $N$ zeros of the hypergeometric
polynomial%
\begin{equation}
P_{N}\left( \alpha _{1},\alpha _{2};\beta _{1};z\right) =\sum_{m=0}^{N}\left[
\frac{\left( -N\right) _{m}\left( \alpha _{1}\right) _{m}~\left( \alpha
_{2}\right) _{m}~z^{N-m}}{m!~\left( \beta _{1}\right) _{m}}\right] ~.
\end{equation}%
Then \textit{Proposition 2.1} implies that these $N$ zeros $\zeta _{n}$
satisfy the following system of $N$ nonlinear algebraic equations%
\begin{eqnarray}
-\alpha _{1}~\alpha _{2}+(N-1)~(\alpha _{1}+\alpha _{2}+1)+\beta _{1}~\zeta
_{n} &&  \notag \\
+2~(3-N+\alpha _{1}+\alpha _{2}-\zeta _{n})~\sigma _{n}^{(1,1)}(\underline{%
\zeta })+3~\sigma _{n}^{(2,2)}(\underline{\zeta })+3~[\sigma _{n}^{(1,1)}(%
\underline{\zeta })]^{2} &=&0~,  \notag \\
n=1,...,N~. &&  \label{EqsForZerosOfPolWithp=q=1}
\end{eqnarray}

As for \textit{Proposition 2.2,} it implies in this case that the following $%
N\times N$ matrix\textbf{\ }$\underline{L}\left( \underline{\zeta }\right) $%
, defined componentwise as follows (see (\ref{Mnmtilde}) with (\ref{fnm}), (%
\ref{gnm}) and (\ref{sigmanpq})), 
\begin{subequations}
\begin{eqnarray}
L_{nn}\left( \underline{\zeta }\right) &=&2~\left[ \frac{\beta _{1}}{2}%
+\left( N-3-\alpha _{1}-\alpha _{2}\right) ~\sigma _{n}^{\left( 1,2\right)
}\left( \underline{\zeta }\right) +\sigma _{n}^{\left( 2,2\right) }\left( 
\underline{\zeta }\right) -3~\sigma _{n}^{\left( 2,3\right) }\left( 
\underline{\zeta }\right) \right.  \notag \\
&&\left. +3~\sigma _{n}^{\left( 1,1\right) }\left( \underline{\zeta }\right)
~\sigma _{n}^{\left( 1,2\right) }\left( \underline{\zeta }\right) \right]
~,~~~n=1,2,...,N~,
\end{eqnarray}%
\begin{eqnarray}
L_{nm}\left( \underline{\zeta }\right) &=&2~\zeta _{n}~\left[ \frac{\alpha
_{1}+\alpha _{2}-N-\zeta _{n}-3~\sigma _{n}^{\left( 1,1\right) }\left( 
\underline{\zeta }\right) }{\left( \zeta _{n}-\zeta _{m}\right) ^{2}}\right.
\notag \\
&&\left. +\frac{3~\zeta _{n}}{\left( \zeta _{n}-\zeta _{m}\right) ^{3}}%
\right] ~,~~~n,m=1,...,N~,~~~n\neq m~,
\end{eqnarray}%
features (again) the $N$\ eigenvalues\textbf{\ }%
\begin{equation}
\lambda _{m}=m~\left( \beta _{1}-1+m\right) ~,~~~m=1,...,N~.
\end{equation}

Note the \textit{isospectral} character of this matrix $\underline{L}\left( 
\underline{\zeta }\right) ,$ which depends explicitly on the $2$ parameters $%
\alpha =\alpha _{1}+\alpha _{2}$ and $\beta _{1}$ and implicitly on the $3$
parameters $\alpha _{1},\alpha _{2}$ and $\beta _{1}$ via the dependence on
these $3$ parameters of the $N$ zeros $\zeta _{n}$ of the polynomial $%
P_{N}\left( \alpha _{1},\alpha _{2};\beta _{1};z\right) $, while its
eigenvalues $\lambda _{m}$ only depend on the single parameter $\beta _{1}$.

\subsection{The case $p=q=2,$ $r=0$}

For $p=q=2,$ $r=0$ (for the definition of $r$ see \textit{Remark 2.5}),
implying (see (\ref{Parama}) and (\ref{Paramb})) 
\end{subequations}
\begin{eqnarray}
a_{0} &=&\alpha _{1}~\alpha _{2}~,~~\ a_{1}=-\left( \alpha _{1}+\alpha
_{2}\right) ~,~~~a_{2}=1~,  \notag \\
b_{1} &=&\left( 1-\beta _{1}\right) ~\left( 1-\beta _{2}\right)
~,~~~b_{2}=2-\beta _{1}-\beta _{2}~,~~\ b_{3}=1~,  \label{abp=q=2}
\end{eqnarray}%
let the $N$ numbers $\zeta _{n}$ be the $N$ zeros of the hypergeometric
polynomial%
\begin{equation}
P_{N}\left( \alpha _{1},\alpha _{2};\beta _{1},\beta _{2};z\right)
=\sum_{m=0}^{N}\left[ \frac{\left( -N\right) _{m}\left( \alpha _{1}\right)
_{m}~\left( \alpha _{2}\right) _{m}~z^{N-m}}{m!~\left( \beta _{1}\right)
_{m}~\left( \beta _{2}\right) _{m}}\right] ~.  \label{PNp=q=2}
\end{equation}%
Then \textit{Proposition 2.1} implies that these $N$ zeros $\zeta _{n}$
satisfy the following system of $N$ nonlinear algebraic equations%
\begin{eqnarray}
-\alpha _{1}\alpha _{2}+(N-1)~(\alpha _{1}+\alpha _{2}+1)+\beta _{1}~\beta
_{2}~\zeta _{n} &&  \notag \\
-2~\left[ (1+\beta _{1}+\beta _{2})\zeta _{n}-\alpha _{1}-\alpha _{2}+N-3%
\right] ~\sigma _{n}^{(1,1)}(\underline{\zeta }) &&  \notag \\
+3~(\zeta _{n}-1)~\left\{ \left[ \sigma _{n}^{(1,1)}(\underline{\zeta })%
\right] ^{2}-\sigma _{n}^{(2,2)}(\underline{\zeta })\right\}
=0~,~~~n=1,...,N~. &&
\end{eqnarray}

As for \textit{Proposition 2.2,} it implies in this case that the following $%
N\times N$ matrix\textbf{\ }$\underline{L}\left( \underline{\zeta }\right) $%
, defined componentwise as follows (see (\ref{Mnmtilde}) with (\ref{fnm}), (%
\ref{gnm}) and (\ref{sigmanpq})), 
\begin{subequations}
\begin{eqnarray}
L_{nn}\left( \underline{\zeta }\right) =\beta _{1}~\beta _{2}+\left[
5+2~\left( \beta _{1}+\beta _{2}\right) \right] ~\sigma _{n}^{\left(
2,2\right) }\left( \underline{\zeta }\right) +6~\sigma _{n}^{\left(
3,3\right) }\left( \underline{\zeta }\right) &&  \notag \\
+2~\left[ \left( N-3-\alpha _{1}-\alpha _{2}\right) ~\sigma _{n}^{\left(
1,2\right) }\left( \underline{\zeta }\right) -3~\sigma _{n}^{\left(
2,3\right) }\left( \underline{\zeta }\right) +3~\sigma _{n}^{\left(
1,1\right) }\left( \underline{\zeta }\right) ~\sigma _{n}^{\left( 1,2\right)
}\left( \underline{\zeta }\right) \right] &&  \notag \\
-3~\sigma _{n}^{\left( 1,1\right) }\left( \underline{\zeta }\right) ~\left[
\sigma _{n}^{\left( 1,1\right) }\left( \underline{\zeta }\right) +2~\sigma
_{n}^{\left( 2,2\right) }\left( \underline{\zeta }\right) \right]
~,~~~n=1,2,...,N~, &&  \label{Ldiagp=q=2}
\end{eqnarray}%
\begin{eqnarray}
L_{nm}\left( \underline{\zeta }\right) &=&2~\zeta _{n}~\left[ \frac{\alpha
_{1}+\alpha _{2}-N+\left( 2-\beta _{1}-\beta _{2}\right) ~\zeta
_{n}+3~\left( \zeta _{n}-1\right) ~\sigma _{n}^{\left( 1,1\right) }\left( 
\underline{\zeta }\right) }{\left( \zeta _{n}-\zeta _{m}\right) ^{2}}\right.
\notag \\
&&\left. -\frac{3~\zeta _{n}~\left( \zeta _{n}-1\right) }{\left( \zeta
_{n}-\zeta _{m}\right) ^{3}}\right] ~,~~~n,m=1,...,N~,~~~n\neq m~,
\label{Lnondiagp=q=2}
\end{eqnarray}%
features the $N$\ eigenvalues\textbf{\ }%
\begin{equation}
\lambda _{m}=m~\left( \beta _{1}-1+m\right) ~\left( \beta _{2}-1+m\right)
~,~~~m=1,...,N~.  \label{landap=q=2}
\end{equation}

Note the isospectral character of this matrix $\underline{L}\left( 
\underline{\zeta }\right) ,$ which depends explicitly on the $3$ parameters $%
\alpha =\alpha _{1}+\alpha _{2}$, $\beta _{1}$ and $\beta _{2}$ and
implicitly on the $4$ parameters $\alpha _{1},\alpha _{2}$, $\beta _{1}$ and 
$\beta _{2}$ via the dependence on these $4$ parameters of the $N$ zeros $%
\zeta _{n}$ of the polynomial $P_{N}\left( \alpha _{1},\alpha _{2};\beta
_{1},\beta _{2};z\right) $, while its eigenvalues $\lambda _{m}$ only depend
on the $2$ parameter $\beta _{1}$ and $\beta _{2}$.

\subsection{The case $p=q=2$, $r=1$}

For $p=q=2$, $r=1$ (for the definition of $r$ see \textit{Remark 2.5}) we
have (as above, see (\ref{abp=q=2}), but now with $\beta _{2}=\alpha _{2}$) 
\end{subequations}
\begin{eqnarray}
a_{0} &=&\alpha _{1}~\alpha _{2}~,~~~a_{1}=-\left( \alpha _{1}+\alpha
_{2}\right) ,~~~a_{2}=1~,  \notag \\
b_{1} &=&\left( 1-\beta _{1}\right) ~\left( 1-\alpha _{2}\right)
~,~~~b_{2}=2-\beta _{1}-\alpha _{2}~,~~\ b_{3}=1~,
\end{eqnarray}%
and we now define the $N$ numbers $\zeta _{n}$ as the $N$ zeros of the
polynomial (\ref{PNp=q=1}) (rather than (\ref{PNp=q=2}); so that these $N$
zeros do not depend on the \textit{arbitrary} parameter $\beta _{2}=\alpha
_{2}$). Then \textit{Proposition 2.1} implies that these $N$ zeros $\zeta
_{n}$ satisfy the following system of $N$ nonlinear algebraic equations 
\begin{subequations}
\begin{eqnarray}
-\alpha _{1}~\alpha _{2}+(N-1)~(\alpha _{1}+\alpha _{2}+1)+\beta _{1}~\alpha
_{2}~\zeta _{n} &&  \notag \\
+2~\left[ 3-N+\alpha _{1}+\alpha _{2}-(1+\beta _{1}+\alpha _{2})~\zeta _{n}%
\right] ~\sigma _{n}^{(1,1)}(\underline{\zeta }) &&  \notag \\
+3(\zeta _{n}-1)\left\{ \left[ \sigma _{n}^{(1,1)}(\underline{\zeta })\right]
^{2}-\sigma _{n}^{(2,2)}(\underline{\zeta })\right\} =0~,~~~n=1,...,N~, &&
\end{eqnarray}%
---which is \textit{different} from (\ref{12c}), although satisfied by the 
\textit{same} zeros. But since this system of $N$ equations must hold for
arbitrary values of the parameters $\alpha _{2},$ it amounts in fact to the
following two separate systems of $N$ equations: 
\end{subequations}
\begin{subequations}
\begin{equation}
-\alpha _{1}+N-1+\beta _{1}~\zeta _{n}+2~\left( 1-\zeta _{n}\right) ~\sigma
_{n}^{(1,1)}(\underline{\zeta })=0~,~~~n=1,...,N~,  \label{Jac1}
\end{equation}%
\begin{eqnarray}
(N-1)~(\alpha _{1}+1)+2~\left[ 3-N+\alpha _{1}-(1+\beta _{1})~\zeta _{n}%
\right] ~\sigma _{n}^{(1,1)}(\underline{\zeta }) &&  \notag \\
+3(\zeta _{n}-1)\left\{ \left[ \sigma _{n}^{(1,1)}(\underline{\zeta })\right]
^{2}-\sigma _{n}^{(2,2)}(\underline{\zeta })\right\} =0~,~~~n=1,...,N~, &&
\label{Jac2}
\end{eqnarray}%
which must both be satisfied by the $N$ zeros $\zeta _{n}$ of the polynomial
(\ref{PNp=q=1}). Indeed the first of these two systems coincides with (\ref%
{12c}); while the second is new (but in fact both these systems of $N$
equations are not quite new, see Subsection 2.5).

As for \textit{Proposition 2.2,} it implies in this case that the matrix $%
\underline{L}\left( \underline{\zeta }\right) $ defined by (\ref{Ldiagp=q=2}%
) and (\ref{Lnondiagp=q=2}) (of course with $\beta _{2}=\alpha _{2}$) hence
reading 
\end{subequations}
\begin{subequations}
\label{LL}
\begin{eqnarray}
L_{nn}\left( \underline{\zeta }\right) =\beta _{1}~\alpha _{2}+\left[
5+2~\left( \beta _{1}+\alpha _{2}\right) \right] ~\sigma _{n}^{\left(
2,2\right) }\left( \underline{\zeta }\right) +6~\sigma _{n}^{\left(
3,3\right) }\left( \underline{\zeta }\right) &&  \notag \\
+2~\left[ \left( N-3-\alpha _{1}-\alpha _{2}\right) ~\sigma _{n}^{\left(
1,2\right) }\left( \underline{\zeta }\right) -3~\sigma _{n}^{\left(
2,3\right) }\left( \underline{\zeta }\right) +3~\sigma _{n}^{\left(
1,1\right) }\left( \underline{\zeta }\right) ~\sigma _{n}^{\left( 1,2\right)
}\left( \underline{\zeta }\right) \right] &&  \notag \\
-3~\sigma _{n}^{\left( 1,1\right) }\left( \underline{\zeta }\right) ~\left[
\sigma _{n}^{\left( 1,1\right) }\left( \underline{\zeta }\right) +2~\sigma
_{n}^{\left( 2,2\right) }\left( \underline{\zeta }\right) \right]
~,~~~n=1,2,...,N~, &&  \label{LLa}
\end{eqnarray}%
\begin{eqnarray}
L_{nm}\left( \underline{\zeta }\right) &=&2~\zeta _{n}~\left[ \frac{\alpha
_{1}+\alpha _{2}-N+\left( 2-\beta _{1}-\alpha _{2}\right) ~\zeta
_{n}+3~\left( \zeta _{n}-1\right) ~\sigma _{n}^{\left( 1,1\right) }\left( 
\underline{\zeta }\right) }{\left( \zeta _{n}-\zeta _{m}\right) ^{2}}\right.
\notag \\
&&\left. -\frac{3~\zeta _{n}~\left( \zeta _{n}-1\right) }{\left( \zeta
_{n}-\zeta _{m}\right) ^{3}}\right] ~,~~~n,m=1,...,N~,~~~n\neq m~,
\label{LLb}
\end{eqnarray}%
---but now with the $N$ zeros $\zeta _{n}$ in these two formulas being again
those of the generalized hypergeometric polynomial (\ref{PNp=q=1}) rather
than (\ref{PNp=q=2}), so that they only depend on $N,$ $\alpha _{1}$ and $%
\beta _{1}$ (but not on $\beta _{2}=\alpha _{2}$)---features the $N$
eigenvalues%
\begin{equation}
\lambda _{m}=m~\left( \beta _{1}-1+m\right) ~\left( \alpha _{2}-1+m\right)
~,~~~m=1,...,N~.  \label{LLc}
\end{equation}%
But again, since these properties must hold for arbitrary values of the
parameter $\alpha _{2},$ they amount to two separate statements, the first
of which is easily seen to reproduce the statement that the matrix (\ref{12d}%
) features the eigenvalues (\ref{12e}), while the second states that the
above matrix (\ref{LLa}), (\ref{LLb}) with $\alpha _{2}=0,$ i. e. 
\end{subequations}
\begin{subequations}
\label{LLL}
\begin{eqnarray}
L_{nn}\left( \underline{\zeta }\right) =\left( 5+2~\beta _{1}\right) ~\sigma
_{n}^{\left( 2,2\right) }\left( \underline{\zeta }\right) +6~\sigma
_{n}^{\left( 3,3\right) }\left( \underline{\zeta }\right) &&  \notag \\
+2~\left[ \left( N-3-\alpha _{1}\right) ~\sigma _{n}^{\left( 1,2\right)
}\left( \underline{\zeta }\right) -3~\sigma _{n}^{\left( 2,3\right) }\left( 
\underline{\zeta }\right) +3~\sigma _{n}^{\left( 1,1\right) }\left( 
\underline{\zeta }\right) ~\sigma _{n}^{\left( 1,2\right) }\left( \underline{%
\zeta }\right) \right] &&  \notag \\
-3~\sigma _{n}^{\left( 1,1\right) }\left( \underline{\zeta }\right) ~\left[
\sigma _{n}^{\left( 1,1\right) }\left( \underline{\zeta }\right) +2~\sigma
_{n}^{\left( 2,2\right) }\left( \underline{\zeta }\right) \right]
~,~~~n=1,2,...,N~, &&  \label{LLLa}
\end{eqnarray}%
\begin{eqnarray}
L_{nm}\left( \underline{\zeta }\right) &=&2~\zeta _{n}~\left[ \frac{\alpha
_{1}-N+\left( 2-\beta _{1}\right) ~\zeta _{n}+3~\left( \zeta _{n}-1\right)
~\sigma _{n}^{\left( 1,1\right) }\left( \underline{\zeta }\right) }{\left(
\zeta _{n}-\zeta _{m}\right) ^{2}}\right.  \notag \\
&&\left. -\frac{3~\zeta _{n}~\left( \zeta _{n}-1\right) }{\left( \zeta
_{n}-\zeta _{m}\right) ^{3}}\right] ~,~~~n,m=1,...,N~,~~~n\neq m~,
\label{LLLb}
\end{eqnarray}%
features the eigenvalues (\ref{LLc}) with $\alpha _{2}=0,$ i. e.%
\begin{equation}
\lambda _{m}=m~\left( m-1\right) \left( \beta _{1}-1+m\right)
~,~~~m=1,...,N~.  \label{LLLc}
\end{equation}

Note that, while these eigenvalues depend on the parameters $\beta _{1}$,
they do \textit{not} depend on the parameter $\alpha _{1}$; hence in this
case the matrix $\underline{L}\left( \underline{\zeta }\right) $, which
itself depends on the 2 parameters $\alpha _{1}$ and $\beta _{1}$, is 
\textit{isospectral} for variations of the parameter $\alpha _{1}.$

\subsection{Results for Jacobi polynomials}

For $p=q=1$---or, equivalently, for $\hat{p}=\hat{q}=1$ (for this notation
see \textit{Remark 2.5})---the generalized hypergeometric polynomial is
simply related to the Jacobi polynomial (see eq. 10.8(16) of \cite{HTF2}):
the transformation (up to an irrelevant multiplicative constant) from the
generalized hypergeometric polynomial $P_{N}\left( \alpha _{1};\beta
_{1};z\right) $, see (\ref{PNp=q=1}), to the standard Jacobi polynomial $%
P_{N}^{\left( \alpha ,\beta \right) }\left( x\right) $ (see \cite{HTF2})
corresponds to the change of variables $\beta _{1}=\alpha +1$, $\alpha
_{1}=N+\alpha +\beta +1$ and $z=2/\left( 1-x\right) $. It can thereby be
verified (with some labor) that the results---corresponding to \textit{%
Proposition 2.1} with $p=q=1$, $r=0$ respectively $p=q=2$, $r=1$---reported
above for these cases reproduce known results \cite{ABCOP1979}: specifically
(\ref{12c}) (or, equivalently, (\ref{Jac1})) respectively (\ref{Jac2})
reproduce (up to appropriate notational changes) eqs. (5.2a) respectively
(5.2b) of \cite{ABCOP1979}. 

On the other hand \textit{Proposition 2.2} with the above assignments of $p,$
$q$ and $r$ seem to produce \textit{new} results for the $N$ zeros $x_{n}$
of the Jacobi polynomial $P_{N}^{\left( \alpha ,\beta \right) }\left(
x\right) $, as displayed below. Indeed for $p=q=1$, $r=0$ it yields the
following 

\textit{Proposition 2.5.1}. The $N\times N$ matrix $\underline{L}\left( 
\underline{x}\right) $ defined componentwise, in terms of the $N$ zeros $%
x_{n}$ of the Jacobi polynomial $P_{N}^{\left( \alpha ,\beta \right) }\left(
x\right) $ and the two parameters $\alpha $ and $\beta $ as follows, 
\end{subequations}
\begin{subequations}
\label{LLLL}
\begin{eqnarray}
&&L_{nm}\left( \underline{x}\right) =\delta _{nm}~\left\{ \alpha
+1+\sum_{\ell =1,~\ell \neq n}^{N}\left[ \frac{\left( 1+x_{\ell }\right)
~\left( 1-x_{n}\right) ^{2}}{\left( x_{n}-x_{\ell }\right) ^{2}}\right]
\right\}   \notag \\
&&-\left( 1-\delta _{nm}\right) ~\left[ \frac{\left( 1+x_{n}\right) ~\left(
1-x_{m}\right) ^{2}}{\left( x_{n}-x_{m}\right) ^{2}}\right] ~,
\end{eqnarray}%
has the $N$ eigenvalues%
\begin{equation}
\lambda _{m}=m~\left( m+\alpha \right) ~,~~~m=1,...,N~.~\square 
\end{equation}

Let us again note the \textit{isospectral} property of this matrix $%
\underline{L}\left( \underline{x}\right) $, whose elements depend, via the $N
$ zeros $x_{n}\equiv x_{n}\left( \alpha ,\beta \right) ,$ on the two
parameters $\alpha $ and $\beta ,$ while its eigenvalues depend only on the
parameter $\alpha $.

Also note that Corollary 5.2.2 of \cite{ABCOP1979} with $s=N-1$, $r=0,$ and
of course $n=N$, identifies an $N\times N$ matrix $G$ (see eq. (5.17) of 
\cite{ABCOP1979} with $C$ and $X$ defined by eq. (5.4) and (5.3) of \cite%
{ABCOP1979}) defined componentwise as follows:
\end{subequations}
\begin{subequations}
\label{GG}
\begin{eqnarray}
G_{nm}=\delta_{nm}\sum_{\ell=1, \ell \neq n}^N\frac{(1-x_\ell^2)(1-x_n)}{(x_n-x_\ell)^2}-(1-\delta_{nm})\frac{(1-x_m)^2(1+x_m)}{(x_n-x_m)^2},
\end{eqnarray}%
and states that its $N$ eigenvalues $g_{m}$ read as follows (see eq. (5.19)
of \cite{ABCOP1979}, with $N-m$ replaced by $m-1$ since these numbers span
the same set of values---from $0$ to $N-1$---for $m=1,...,N$): 
\begin{equation}
g_{m}=\left( m-1\right) ~\left( m+\alpha -1\right) ~,~~\ m=1,...,N~.
\end{equation}%
These formulas, (\ref{GG}), are similar to, but different from, (\ref{LLLL}%
); although of course there must be a way to relate them, since they hold
for the same set of $N+1$ numbers $\alpha $ and $x_{n}\equiv x_{n}\left(
\alpha ,\beta \right) $. 

And likewise for $p=q=2$, $r=1$ it yields---via a treatment analogous to
that of Section 2.4---the following

\textit{Proposition 2.5.2}. The $N\times N$ matrix $\underline{L}\left( 
\underline{x}\right) $ defined componentwise, in terms of the $N$ zeros $%
x_{n}$ of the Jacobi polynomial $P_{N}^{\left( \alpha ,\beta \right) }\left(
x\right) $ and the two parameters $\alpha $ and $\beta $ as follows, 
\end{subequations}
\begin{subequations}
\begin{eqnarray}
L_{nn} &=&\left( 7+2\alpha \right) ~\sigma _{n}^{\left( 2,2\right) }\left( 
\underline{x}\right) +6~\sigma _{n}^{\left( 3,3\right) }\left( \underline{x}%
\right) -2~\left( 4+\alpha +\beta \right) ~\sigma _{n}^{\left( 1,2\right)
}\left( \underline{x}\right) -6~\sigma _{n}^{\left( 2,3\right) }\left( 
\underline{x}\right)   \notag \\
&&-3~\left[ \sigma _{n}^{\left( 1,1\right) }\left( \underline{x}\right) %
\right] ^{2}-6~\sigma _{n}^{\left( 1,1\right) }\left( \underline{x}\right)
~\sigma _{n}^{\left( 2,2\right) }\left( \underline{x}\right) +6~\sigma_n^{\left(1,1\right)}\left(\underline{x}\right)~\sigma_n^{\left(1,2\right)}\left(\underline{x}\right)~,
\end{eqnarray}%
\begin{eqnarray}
L_{nm} &=&\left[ \left( \alpha +\beta +1\right) ~\left( 1-x_{n}\right)
+2~\left( 1-\alpha \right) +3~\left( 1+x_{n}\right) ~\sigma _{n}^{\left(
1,1\right) }\left( \underline{x}\right) \right] ~\left( \frac{1-x_{m}}{%
x_{n}-x_{m}}\right) ^{2}  \notag \\
&&-3~\left( 1+x_{n}\right) ~\left( \frac{1-x_{m}}{x_{n}-x_{m}}\right) ^{3}~,
\end{eqnarray}%
where (see (\ref{sigmanpq})) 
\begin{equation}
\sigma _{n}^{\left( r,\rho \right) }\left( \underline{x}\right) ==\sum_{\ell
=1;\ell \neq n}^{N}\left[ \left( \frac{2}{1-x_{\ell }}\right) ^{r-\rho
}~\left( \frac{1-x_{n}}{x_{n}-x_{\ell }}\right) ^{\rho }\right] ~,
\end{equation}%
has the $N$ eigenvalues%
\begin{equation}
\lambda _{m}=m~\left( m-1\right) ~\left( m+\alpha \right)
~,~~~m=1,...,N~.~\square 
\end{equation}%
Note again the \textit{isospectral} character of this matrix, which depends
on the two parameters $\alpha $ and $\beta $, while its eigenvalues depend
only on the parameter $\alpha $. 

\section{Proofs}

In this section we prove the findings reported in the preceding Section 2.

Let the $t$-dependent monic polynomial, of degree $N$ in $z$ and
characterized by its $N$ zeros $z_{n}\left( t\right) $ and its $N$
coefficients $c_{m}\left( t\right) $, 
\end{subequations}
\begin{eqnarray}
\psi _{N}\left( z,t\right) &=&\tprod\limits_{n=1}^{N}\left[ z-z_{n}\left(
t\right) \right] =z^{N}+\sum_{m=1}^{N}\left[ c_{m}\left( t\right) ~z^{N-m}%
\right]  \notag \\
&=&\sum_{m=0}^{N}\left[ c_{m}\left( t\right) ~z^{N-m}\right] ~~~\text{with~~~%
}c_{0}=1~,  \label{psizt}
\end{eqnarray}%
satisfy the \textit{linear} Partial Differential Equation (PDE)%
\begin{eqnarray}
&&\left( \frac{\partial }{\partial ~t}\right) ~\psi _{N}\left( z,t\right)
=-\left\{ \left( z~\frac{\partial }{\partial ~z}-N\right)
\tprod\limits_{j=1}^{q}\left[ \beta _{j}-1-\left( z~\frac{\partial }{%
\partial ~z}-N\right) \right] \right.  \notag \\
&&\left. -\left( \frac{\partial }{\partial ~z}\right) \tprod\limits_{j=1}^{p}%
\left[ \alpha _{j}-\left( z~\frac{\partial }{\partial ~z}-N\right) \right]
\right\} ~\psi _{N}\left( z,t\right) ~.  \label{PDE}
\end{eqnarray}%
Clearly this implies that its coefficients $c_{m}\left( t\right) $ satisfy
the following system of $N$ \textit{linear} Ordinary Differential Equations
(ODEs): 
\begin{subequations}
\label{EqMotcm}
\begin{eqnarray}
\left( \frac{d}{d~t}\right) ~c_{m}\left( t\right) =m~\left[
\tprod\limits_{j=1}^{q}\left( \beta _{j}-1+m\right) \right] ~c_{m}\left(
t\right) &&  \notag \\
+\left( N+1-m\right) ~\left[ \tprod\limits_{j=1}^{p}\left( \alpha
_{j}-1+m\right) \right] ~c_{m-1}\left( t\right) ~,~~~m=1,...,N~, &&
\label{ODEs}
\end{eqnarray}%
with the conditions (see (\ref{psizt}))%
\begin{equation}
c_{0}\left( t\right) =1~,  \label{Condca}
\end{equation}%
\begin{equation}
c_{j}\left( t\right) =0~~~\text{for~~~}j<0~~~\text{and for~~~}j>N~.
\label{Condcb}
\end{equation}%
Hence the general solution of this system of $N$ \textit{linear} ODEs reads
as follows: 
\end{subequations}
\begin{equation}
\underline{c}\left( t\right) =\sum_{m=1}^{N}\left[ \tilde{\eta}_{m}~\exp
\left( \tilde{\lambda}_{m}~t\right) ~\underline{\tilde{v}}^{\left( m\right) }%
\right] ~,  \label{Solc}
\end{equation}%
where the $N$ ($t$-independent) parameters $\tilde{\eta}_{m}$ can be
arbitrarily assigned (or adjusted to satisfy the $N$ \textit{initial}
conditions $c_{m}\left( 0\right) $), while the numbers $\tilde{\lambda}_{m}$
respectively the $N$-vectors $\underline{\tilde{v}}^{\left( m\right) }$ are
clearly the $N$ eigenvalues respectively the $N$ eigenvectors of the
algebraic eigenvalue problem 
\begin{subequations}
\label{M}
\begin{equation}
\underline{\Lambda }~\underline{\tilde{v}}^{\left( m\right) }=\tilde{\lambda}%
_{m}~\underline{\tilde{v}}^{\left( m\right) }~,~~~m=1,...,N~,
\end{equation}%
with the $N\times N$ matrix $\underline{\Lambda }$ defined componentwise as
follows:%
\begin{equation}
\Lambda _{m,m}=m~\tprod\limits_{j=1}^{q}\left( \beta _{j}-1+m\right)
~,~~~m=1,...,N~,  \label{Ma}
\end{equation}%
\begin{equation}
\Lambda _{m,m-1}=\left( N+1-m\right) ~\tprod\limits_{j=1}^{p}\left( \alpha
_{j}-1+m\right) ~,~~~m=2,...,N~,  \label{Mb}
\end{equation}%
with all other elements vanishing, $\Lambda _{m,n}=0$ unless $n=m$ or $n=m-1$%
.

The (lower) \textit{triangular} character of the matrix $\underline{\Lambda }
$ implies that its $N$ eigenvalues $\tilde{\lambda}_{m}$ can be explicitly
evaluated: 
\end{subequations}
\begin{equation}
\tilde{\lambda}_{m}=\Lambda _{m,m}=m~\tprod\limits_{j=1}^{q}\left( \beta
_{j}-1+m\right) ~,~~~m=1,...,N~.  \label{mum}
\end{equation}

Let us now denote as $\bar{\psi}_{N}\left( z\right) $ the $t$-independent 
\textit{monic polynomial} solution of (\ref{PDE}),%
\begin{eqnarray}
\bar{\psi}_{N}\left( z\right) &=&\tprod\limits_{n=1}^{N}\left[ z-\zeta _{n}%
\right] =z^{N}+\sum_{m=1}^{N}\left[ \gamma _{m}~z^{N-m}\right]  \notag \\
&=&\sum_{m=0}^{N}\left[ \gamma _{m}~z^{N-m}\right] ~~~\text{with~~~}\gamma
_{0}=1~,  \label{psibar}
\end{eqnarray}%
hence the monic polynomial solution of the ODE%
\begin{eqnarray}
&&\left\{ \left( z~\frac{d}{d~z}-N\right) ~\tprod\limits_{j=1}^{q}\left[
\beta _{j}-1-\left( z~\frac{d}{d~z}-N\right) \right] \right.  \notag \\
&&\left. -\left( \frac{d}{d~z}\right) ~\tprod\limits_{j=1}^{p}\left[ \alpha
_{j}-\left( z~\frac{d}{d~z}-N\right) \right] \right\} ~\bar{\psi}_{N}\left(
z\right) =0~.  \label{Eqpsibar}
\end{eqnarray}%
Note that we denote as $\gamma _{m}$ its coefficients and as $\zeta _{n}$
its $N$ zeros, see (\ref{psibar}) (the fact that the notation for the zeros
is identical to that used above, see (\ref{zeros}), is not accidental: see
below). Clearly the $t$-independent coefficients $\gamma _{m}$ correspond to
the "equilibrium configuration" $c_{m}\left( t\right) =\gamma _{m}$ of the
linear "dynamical system" (\ref{ODEs}), hence they are characterized as the
solutions of the system of $N$ \textit{linear algebraic} equations 
\begin{subequations}
\label{Eqgammas}
\begin{equation}
\left( m+1\right) ~\left[ \tprod\limits_{j=1}^{q}\left( \beta _{j}+m\right) %
\right] ~\gamma _{m+1}=\left( m-N\right) ~\left[ \tprod\limits_{j=1}^{p}%
\left( \alpha _{j}+m\right) \right] ~\gamma _{m}~,~~~m=0,...,N~,
\label{Eqgamma}
\end{equation}%
with the conditions 
\begin{equation}
\gamma _{0}=1~,~~~\gamma _{N+1}=0~.  \label{BoundCondgamma}
\end{equation}%
The first of these conditions corresponds to (\ref{Condca}); the second
(which via the recursion (\ref{Eqgamma}) clearly implies $\gamma _{m}=0$ for 
$m>N$) corresponds to (\ref{Condcb}) and it is automatically satisfied
because the right-hand side of the recursion (\ref{Eqgamma}) vanishes for $%
m=N$ due to the factor $\left( m-N\right) $ (and note that no conditions
need to be assigned on $\gamma _{m}$ with $m<0$ since no such values enter
in the recursion (\ref{Eqgamma}); in any case for $m=-1$ the recursion (\ref%
{Eqgamma}) would imply $\gamma _{-1}=0$ since its left-hand side vanishes
due to the factor $\left( m+1\right) $).

It is then plain from the two-term (hence explicitly solvable) recursion
relation (\ref{Eqgammas}) that the $N+1$ parameters $\gamma _{m}$ read as
follows: 
\end{subequations}
\begin{equation}
\gamma _{m}=\frac{\left( -N\right) _{m}~\tprod\limits_{j=1}^{p}\left( \alpha
_{j}\right) _{m}}{m!~\tprod\limits_{j=1}^{q}\left( \beta _{j}\right) _{m}}%
,~~~m=0,1,...,N~.  \label{gammas}
\end{equation}

This---besides implying that the ODE (\ref{Eqpsibar}) does possess a
polynomial solution $\bar{\psi}\left( z\right) $ of degree $N$ in $z$%
---shows that this $t$-independent polynomial solution $\bar{\psi}\left(
z\right) $ of the PDE (\ref{PDE}) coincides with the \textit{generalized
hypergeometric polynomial} (\ref{PN}) (compare (\ref{PN}) to (\ref{psibar})
with (\ref{gammas})); of course up to an overall multiplicative constant,
which can be arbitrarily assigned due to the \textit{linear} character of
the ODE (\ref{Eqpsibar}), and was chosen above so that the polynomial $\bar{%
\psi}\left( z\right) $ be \textit{monic}, see (\ref{psibar}), hence indeed
coincide with the \textit{generalized hypergeometric polynomial} (\ref{PN}).

Our next task is to identify the "equations of motion" characterizing the $t$%
-evolution of the $N$ zeros $z_{n}\left( t\right) $ of $\psi \left(
z,t\right) ,$ see (\ref{psizt}), implied by the PDE (\ref{PDE}) and by the
corresponding $t$-evolution of the $N$ coefficients $c_{m}\left( t\right) ,$
see (\ref{EqMotcm}). To this end it is convenient to firstly reformulate the
PDE (\ref{PDE}) as follows:%
\begin{eqnarray}
\left( \frac{\partial }{\partial ~t}\right) ~\psi _{N}\left( z,t\right)
=-\left\{ \sum_{k=1}^{q+1}\left[ b_{k}~\left( z~\frac{\partial }{\partial ~z}%
-N\right) ^{k}\right] \right.  
\left. -\left( \frac{\partial }{\partial ~z}\right) ~\sum_{j=0}^{p}\left[
a_{j}~\left( z~\frac{\partial }{\partial ~z}-N\right) ^{j}\right] \right\}
~\psi _{N}\left( z,t\right) ~,  \label{PDEab}
\end{eqnarray}%
where of course the new parameters $b_{k}$ respectively $a_{j}$ are related
to the parameters $\beta _{k}$ respectively $\alpha _{j}$ as detailed above,
see (\ref{Paramb}) and (\ref{Parama}).

Then it is easily seen that the equations of motion characterizing the $t$%
-evolution of the $N$ zeros $z_{n}\left( t\right) $ of $\psi \left(
z,t\right) $, see (\ref{psizt}), read as follows (of course below a
superimposed dot denotes a $t$-differentiation): 
\begin{equation}
\dot{z}_{n}=\sum_{k=1}^{q+1}\left[ b_{k}~f_{n}^{\left( k\right) }\left( 
\underline{z}\right) \right] -\sum_{j=0}^{p}\left[ a_{j}~g_{n}^{\left(
j\right) }\left( \underline{z}\right) \right] ~,~~~n=1,...,N~.  \label{Eqznt}
\end{equation}%
Indeed, this clearly follows from the PDE (\ref{PDEab}) via the identities (%
\ref{Iden1}) and (\ref{Iden2}) and the analogous identity corresponding to
the logarithmic $t$-derivative of (\ref{psizt}) hence reading (via the
short-hand notation (\ref{ShortHandNotat}))%
\begin{equation}
\left( \frac{\partial }{\partial ~t}\right) ~\psi _{N}\left( z,t\right)
\Leftrightarrow -\dot{z}_{n}~.
\end{equation}

It is moreover plain from the developments reported above (see (\ref{psibar}%
) and the sentence following (\ref{gammas})) that the coordinates $\zeta
_{n} $ characterizing the \textit{equilibrium} configuration $\underline{%
\zeta }\equiv \left( \zeta _{1},...,\zeta _{n}\right) $ of this system,
---which of course (see (\ref{Eqznt})) satisfy the set of $N$ algebraic
(generally \textit{nonlinear}) equations (\ref{Eqzitan})---coincides with
the $N$ zeros of the generalized hypergeometric polynomial $P_{N}\left(
\alpha _{1},...,\alpha _{p};\beta _{1},...,\beta _{q};z\right) $, see (\ref%
{zeros}).

\textit{Proposition 2.1} is thereby proven.

Our next step is to consider the behavior of the dynamical system (\ref%
{Eqznt}) in the infinitesimal vicinity of its equilibrium configuration $%
\underline{z}\left( t\right) =\underline{\zeta }$. To this end we set%
\begin{equation}
\underline{z}\left( t\right) =\underline{\zeta }+\varepsilon ~\underline{x}%
\left( t\right) ~;~~~z_{n}\left( t\right) =\zeta _{n}+\varepsilon
~x_{n}\left( t\right) ~,~~~n=1,..,N~,
\end{equation}%
with $\varepsilon $ infinitesimal. We thereby \textit{linearize} the
equations of motion (\ref{Eqznt}), getting%
\begin{equation}
\underline{\dot{x}}=\underline{L}~\underline{x}~;~~~\dot{x}%
_{n}=\sum_{m=1}^{N}\left( L_{nm}~x_{m}\right) ~,~~~n=1,...,N~,  \label{Eqxn}
\end{equation}%
with the $N\times N$ matrix $\underline{L}$ defined componentwise by (\ref%
{Mnmtilde}).

The \textit{general} solution of the system of \textit{linear} ODEs (\ref%
{Eqxn}) reads of course then as follows: 
\begin{equation}
\underline{x}\left( t\right) =\sum_{m=1}^{N}\left[ \eta _{m}~\exp \left(
\lambda _{m}~t\right) ~\underline{v}^{\left( m\right) }\right] ~,
\end{equation}%
where the $N$ ($t$-independent) parameters $\eta _{m}$ can be arbitrarily
assigned (or adjusted to satisfy the $N$ \textit{initial} conditions $%
x_{n}\left( 0\right) $), while the numbers $\lambda _{m}$ respectively the $%
N $-vectors $\underline{v}^{\left( m\right) }$ are clearly the $N$
eigenvalues respectively the $N$ eigenvectors of the algebraic eigenvalue
problem (\ref{EigenEqMtilde}). But the behavior of the dynamical system (\ref%
{Eqznt}) in the \textit{immediate vicinity} of its \textit{equilibria}
cannot differ from its \textit{general} behavior, which is characterized by
the $N$ exponentials $\exp \left( \tilde{\lambda}_{m}~t\right) $, as implied
by the relation between the $N$ zeros $z_{n}\left( t\right) $ and the
coefficients $c_{m}\left( t\right) $ of the monic polynomial (of degree $N$
in $z$) $\psi _{N}\left( z,t\right) $, see (\ref{psizt}), and by the
explicit formula (\ref{Solc}) with (\ref{mum}) detailing the time evolution
of the $N$ coefficients $c_{m}\left( t\right) .$ Hence the (set of)
eigenvalues $\lambda _{m}$ of the matrix $\underline{L}$, see (\ref{Mnmtilde}%
), must coincide with the (set of) eigenvalues $\tilde{\lambda}_{m}$, see (%
\ref{mum}), of the matrix $\underline{\Lambda }$, see (\ref{M}).

\textit{Proposition 2.2} is thereby proven.

\bigskip

\section{Outlook}

A follow-up to the present paper shall take advantage of the new polynomial
identities reported in Appendix A in order to identify \textit{new} classes
of \textit{solvable} $N$-body problems of "goldfish" type \cite{Gold} \cite%
{C1978} \cite{C2001} \cite{C2008} involving \textit{many-body} interactions
and featuring several free parameters, by extending in a fairly obvious
manner (indeed, see (\ref{Eqznt})) the approach and findings reported in 
\cite{BC2013}.

Two possible directions of further investigation shall try and extend the
approach and findings, reported in this paper for the $N$\ zeros of
hypergeometric polynomials of order $N$, to the $N$\ zeros of basic
hypergeometric polynomials of order $N$\ and to the, generally infinite,
zeros of (nonpolynomial) generalized hypergeometric functions.

\bigskip

\section{Acknowledgements}

One of us (OB) would like to acknowledge with thanks the hospitality of the
Physics Department of the University of Rome "La Sapienza" on the occasion
of two two-week visits there in June 2012 and May 2013, and the financial
support for these trips provided by the NSF-AWM Travel Grant. The other one
(FC) would like to acknowledge with thanks the hospitality of Concordia
College for a one-week visit there in November 2013, during which time this
paper was essentially finalized.

\bigskip

\section{Appendix A: Polynomial identities}

Several \textit{polynomial identities} are reported in Appendix A of \cite%
{C2008} (see in particular the paperback version of this monograph, where
the identities (A.8k) and (A.8l) are corrected) and in Appendix A of \cite%
{BC2013}. In this appendix we report, and then prove, two additional classes
of analogous identities. These findings have an interest of their own: for
instance they open the way to the identification of new classes of solvable $%
N$-body problems of "goldfish" type (as mentioned in Section 4).

To make this Appendix self-consistent we firstly introduce the notation we
use below, even if this might entail a bit of repetition (see \cite{C2008}).

Let $\psi \left( z\right) $ be a \textit{monic} polynomial of degree $N$ in $%
z$, and denote as $z_{n}$ its $N$ zeros, 
\begin{equation}
\psi \left( z\right) =\tprod\limits_{n=1}^{N}\left( z-z_{n}\right) ~,
\label{psi}
\end{equation}%
and as $\underline{z}$ the $N$-vector of components $z_{n}$, $\underline{z}%
\equiv \left( z_{1},...,z_{N}\right) $.

We then introduce the following short-hand notation: the formula 
\begin{subequations}
\label{ShortHandNotat}
\begin{equation}
D~\psi \left( z\right) \Longleftrightarrow f_{n}\left( \underline{z}\right)
\end{equation}%
stands for the identity%
\begin{equation}
D~\psi \left( z\right) =\sum_{n=1}^{N}\left[ \left( z-z_{n}\right)
^{-1}~f_{n}\left( \underline{z}\right) \right] ~.
\end{equation}%
Here $D$ is a differential operator with polynomial coefficients acting on
the variable $z$, say%
\begin{equation}
D=\sum_{j=0}^{J}\left[ p^{\left( j\right) }\left( z\right) ~\left( \frac{d}{%
d~z}\right) \right] ~,
\end{equation}%
with the coefficients $p^{\left( j\right) }\left( z\right) $ polynomial in $%
z $. The simpler examples of such identities are 
\end{subequations}
\begin{subequations}
\begin{equation}
\left( \frac{d}{d~z}\right) ~\psi \left( z\right) \Longleftrightarrow 1~,
\label{A.4}
\end{equation}

\begin{equation}
\left( z~\frac{d}{d~z}-N\right) ~\psi \left( z\right) \Longleftrightarrow
z_{n}~;  \label{A.6a}
\end{equation}%
clearly the first obtains by logarithmic differentiation of (\ref{psi}), and
the second from the first via the identity $z/\left( z-z_{n}\right)
=1+z_{n}/\left( z-z_{n}\right) $ (these identities, (\ref{A.4}) and \ref%
{A.6a}), are reported in \cite{C2008} as (A.4) and (A.6a)).

Finally, to write more compactly some of the formulas obtained below and
reported in Section 2 we introduce the following convenient notations: 
\end{subequations}
\begin{subequations}
\label{Sums}
\begin{equation}
\sum_{\ell =1;(n)}^{N}\equiv \sum_{\ell =1;\ell \neq n}^{N}~,
\end{equation}%
\begin{equation}
\sum_{\ell ,m=1;(n)}^{N}\equiv \sum_{\ell ,m=1;\ell \neq n,m\neq n,\ell \neq
m}^{N}~,  \label{Sum1}
\end{equation}%
\begin{equation}
\sum_{\ell ,k,m=1;\left( n\right) }^{N}\equiv \sum_{\ell ,k,m=1;\ell \neq
n,k\neq n,m\neq n,\ell \neq k,\ell \neq m,k\neq m}^{N}~;  \label{Sum2}
\end{equation}%
\end{subequations}
\begin{subequations}
\label{dd}
\begin{equation}
d_{n\left( \ell m\right) }^{\left( 3\right) }\left( \underline{z}\right)
=\left( z_{n}-z_{\ell }\right) ~\left( z_{n}-z_{m}\right) ~\left( z_{\ell
}-z_{m}\right) ~,  \label{d3}
\end{equation}%
\begin{equation}
d_{n\left( \ell km\right) }^{\left( 4\right) }\left( \underline{z}\right)
=\left( z_{n}-z_{\ell }\right) ~\left( z_{n}-z_{k}\right) ~\left(
z_{n}-z_{m}\right) ~\left( z_{\ell }-z_{m}\right) ~\left( z_{\ell
}-z_{k}\right) ~\left( z_{k}-z_{m}\right) ~.  \label{d4}
\end{equation}

\textit{Remark A.1}. The upper label on the quantities $d_{n\left( \ell
m\right) }^{\left( 3\right) }\left( \underline{z}\right) $\ and $d_{n\left(
\ell km\right) }^{\left( 4\right) }\left( \underline{z}\right) $\ is a
reminder that $d_{n\left( \ell m\right) }^{\left( 3\right) }\left( 
\underline{z}\right) $\ involves the coordinates of the $3$\ different zeros 
$z_{n},$\ $z_{\ell },$\ $z_{m}$, and $d_{n\left( \ell km\right) }^{\left(
4\right) }\left( \underline{z}\right) $\ involves the coordinates of the $4$%
\ different zeros $z_{n},$\ $z_{\ell },$\ $z_{k},$\ $z_{m}$. It is moreover
plain that these quantities---that shall play the role of denominators in
sums over the zeros, see below---are antisymmetric under the exchange of any
pair of these zeros: this entails that, by taking advantage of these
(anti)symmetries, the arguments of the sums reported below could be
rewritten in many different ways by appropriate exchanges of the dummy
indices being summed upon. We use this property to also obtain convenient
expressions based on the quantities $\sigma _{n}^{\left( r,\rho \right)
}\left( \underline{z}\right) $, introduced immediately below. $\square $

It is now convenient to introduce the quantities 
\end{subequations}
\begin{equation}
\sigma _{n}^{\left( r,\rho \right) }\left( \underline{z}\right) =\sum_{\ell
=1;\ell \neq n}^{N}\left[ \frac{z_{\ell }^{r}}{\left( z_{n}-z_{\ell }\right)
^{\rho }}\right] ~.  \label{sigmanpq}
\end{equation}%
Here and hereafter $r$ and $\rho $ are \textit{arbitrary nonnegative integers%
} (unless otherwise indicated). Clearly this definition implies 
\begin{subequations}
\label{sigmanm}
\begin{equation}
\frac{d~\sigma _{n}^{\left( r,\rho \right) }\left( \underline{z}\right) }{%
d~z_{m}}=-\rho ~\delta _{nm}~\sigma _{n}^{\left( r,\rho +1\right) }\left( 
\underline{z}\right) +\left( 1-\delta _{nm}\right) ~\frac{\left[
r~z_{n}+\left( \rho -r\right) ~z_{m}\right] ~z_{m}^{r-1}}{\left(
z_{n}-z_{m}\right) ^{\rho +1}}
\end{equation}%
hence%
\begin{equation}
\frac{d~\sigma _{n}^{\left( r,r\right) }\left( \underline{z}\right) }{d~z_{m}%
}=-r~\delta _{nm}~\sigma _{n}^{\left( r,r+1\right) }\left( \underline{z}%
\right) +\left( 1-\delta _{nm}\right) ~\frac{r~z_{n}~z_{m}^{r-1}~}{\left(
z_{n}-z_{m}\right) ^{r+1}}~;
\end{equation}%
and it is also obvious (and used below) that 
\end{subequations}
\begin{equation}
\sigma _{n}^{\left( 0,0\right) }\left( \underline{z}\right) =N-1~.
\label{sigma00}
\end{equation}

\textit{Remark A.2}. The quantities $\sigma _{n}^{\left( r,\rho \right)
}\left( \underline{z}\right) ,$ see (\ref{sigmanpq}),\ are related to each
other by trivial identities, for instance clearly the replacement of $%
z_{\ell }^{r}$\ with $z_{\ell }^{r-1}\left[ z_{n}+\left( z_{\ell
}-z_{n}\right) \right] $\ in the numerator of their definition (\ref%
{sigmanpq}) implies 
\begin{equation}
\sigma _{n}^{\left( r,\rho \right) }\left( \underline{z}\right)
=z_{n}~\sigma _{n}^{\left( r-1,\rho \right) }\left( \underline{z}\right)
-\sigma _{n}^{\left( r-1,\rho -1\right) }\left( \underline{z}\right)
~,~~~r,\rho =1,2,3,...  \label{znsigma}
\end{equation}%
yielding (by iteration) 
\begin{subequations}
\begin{equation}
\sigma _{n}^{\left( r,\rho \right) }\left( \underline{z}\right)
=\sum_{k=0}^{s}\left[ \left( -1\right) ^{k}~\binom{s}{k}~z_{n}^{s-k}~\sigma
_{n}^{\left( r-s,\rho -k\right) }\left( \underline{z}\right) \right]
~,~~~s=0,1,2,...,\min \left( r,\rho \right) ~,
\end{equation}%
\ hence in particular%
\begin{equation}
\sigma _{n}^{\left( r,\rho \right) }\left( \underline{z}\right)
=\sum_{k=0}^{r}\left[ \left( -1\right) ^{k}~\binom{r}{k}~z_{n}^{r-k}~\sigma
_{n}^{\left( 0,\rho -k\right) }\left( \underline{z}\right) \right] ~\ ~\text{%
if}~\ ~r\leq \rho ~,
\end{equation}%
\begin{equation}
\sigma _{n}^{\left( r,\rho \right) }\left( \underline{z}\right)
=\sum_{k=0}^{\rho }\left[ \left( -1\right) ^{k}~\binom{\rho }{k}~z_{n}^{\rho
-k}~\sigma _{n}^{\left( r-\rho ,\rho -k\right) }\left( \underline{z}\right) %
\right] ~\ ~\text{if}~\ ~r\geq \rho ~.~\square 
\end{equation}

To obtain the second version of the identities reported below we also used
the identity 
\end{subequations}
\begin{equation}
z_{n}~\sigma _{n}^{\left( r,\rho \right) }\left( \underline{z}\right)
=\sigma _{n}^{\left( r+1,\rho \right) }\left( \underline{z}\right) +\sigma
_{n}^{\left( r,\rho -1\right) }\left( \underline{z}\right)
~,~~~r=0,1,2,...,~\rho =1,2,3,...  \label{znsigman}
\end{equation}%
(which correspond clearly to (\ref{znsigma})), and the property (\ref%
{sigma00}).

The first class of new identities reads as follows: 
\begin{subequations}
\label{Identities1}
\begin{equation}
\left( z~\frac{d}{d~z}-N\right) ^{j}~\psi \left( z\right)
\Longleftrightarrow f_{n}^{\left( j\right) }\left( \underline{z}\right)
~,~~~j=1,2,...~,  \label{Iden1}
\end{equation}%
with the $N$ quantities $f_{n}^{\left( j\right) }\left( \underline{z}\right) 
$ defined recursively as follows:%
\begin{equation}
f_{n}^{\left( j+1\right) }\left( \underline{z}\right) =-f_{n}^{\left(
j\right) }\left( \underline{z}\right) +\sum_{\ell =1;(n)}^{N}\left[ \frac{%
z_{n}~f_{\ell }^{\left( j\right) }\left( \underline{z}\right) +z_{\ell
}~f_{n}^{\left( j\right) }\left( \underline{z}\right) }{z_{n}-z_{\ell }}%
\right] ~,~~~j=1,2,...~,  \label{Recfn}
\end{equation}%
with 
\end{subequations}
\begin{subequations}
\label{ff}
\begin{equation}
f_{n}^{\left( 1\right) }\left( \underline{z}\right) =z_{n}~,  \label{fn1}
\end{equation}%
implying%
\begin{eqnarray}
&&f_{n}^{\left( 2\right) }\left( \underline{z}\right) =-z_{n}+2\sum_{\ell
=1;(n)}^{N}\left( \frac{z_{n}~z_{\ell }}{z_{n}-z_{\ell }}\right) ~,  \notag
\\
&=&z_{n}~\left[ -1+2~\sigma _{n}^{\left( 1,1\right) }\left( \underline{z}%
\right) \right] ~,  \label{fn2}
\end{eqnarray}%
\begin{eqnarray}
&&f_{n}^{\left( 3\right) }\left( \underline{z}\right) =z_{n}-6~\sum_{\ell
=1;(n)}^{N}\left( \frac{z_{n}~z_{\ell }}{z_{n}-z_{\ell }}\right)
+6~\sum_{\ell ,m=1;(n)}^{N}\left( \frac{z_{n}~z_{\ell }^{2}~z_{m}}{%
d_{n\left( \ell m\right) }^{\left( 3\right) }\left( \underline{z}\right) }%
\right)  \notag \\
&=&z_{n}\left\{ 1-6~\sigma _{n}^{\left( 1,1\right) }\left( \underline{z}%
\right) -3~\sigma _{n}^{\left( 2,2\right) }\left( \underline{z}\right) +3~%
\left[ \sigma _{n}^{\left( 1,1\right) }\left( \underline{z}\right) \right]
^{2}\right\} ~,  \label{fn3}
\end{eqnarray}%
\begin{eqnarray}
&&f_{n}^{\left( 4\right) }\left( \underline{z}\right) =-z_{n}+14~\sum_{\ell
=1;(n)}^{N}\left( \frac{z_{n}~z_{\ell }}{z_{n}-z_{\ell }}\right)
-36~\sum_{\ell ,m=1;(n)}^{N}\left( \frac{z_{n}~z_{\ell }^{2}~z_{m}}{d_{n\ell
m}^{\left( 3\right) }\left( \underline{z}\right) }\right)  \notag \\
&&+24~\sum_{\ell ,k,m=1;(n)}^{N}\left( \frac{z_{n}~z_{\ell
}~z_{k}^{3}~z_{m}^{2}~}{d_{n\ell km}^{\left( 4\right) }\left( \underline{z}%
\right) }\right)  \notag \\
&=&z_{n}~\left\{ -1+14~\sigma _{n}^{(1,1)}(\underline{z})+18~\sigma
_{n}^{(2,2)}(\underline{z})+8~\sigma _{n}^{(3,3)}(\underline{z})\right. 
\notag \\
&&\left. -18~\left[ \sigma _{n}^{(1,1)}(\underline{z})\right] ^{2}-12~\sigma
_{n}^{(1,1)}(\underline{z})~\sigma _{n}^{(2,2)}(\underline{z})+4~\left[
\sigma _{n}^{(1,1)}(\underline{z})\right] ^{3}\right\} ~,
\end{eqnarray}%
and so on (via standard, but increasingly tedious, computations).

The second class of new identities reads as follows: 
\end{subequations}
\begin{subequations}
\label{Identities2}
\begin{equation}
\left( \frac{d}{d~z}\right) ~\left( z~\frac{d}{d~z}-N\right) ^{j}~\psi
\left( z\right) \Longleftrightarrow g_{n}^{\left( j\right) }\left( 
\underline{z}\right) ~,~~~j=1,2,...  \label{Iden2}
\end{equation}%
(for the case $j=0$ see (\ref{A.4}) implying $g_{n}^{\left( 0\right) }\left( 
\underline{z}\right) =1$), with the $N$ quantities $g_{n}^{\left( j\right)
}\left( \underline{z}\right) $ defined as follows:%
\begin{equation}
g_{n}^{\left( j\right) }\left( \underline{z}\right) =\sum_{\ell =1;(n)}^{N}%
\left[ \frac{f_{n}^{\left( j\right) }\left( \underline{z}\right) +f_{\ell
}^{\left( j\right) }\left( \underline{z}\right) }{z_{n}-z_{\ell }}\right]
~,~~~j=1,2,...~,  \label{gn}
\end{equation}%
implying (for the second version of these formulas via (\ref{sigma00})) 
\end{subequations}
\begin{subequations}
\label{ggg}
\begin{eqnarray}
&&g_{n}^{\left( 1\right) }\left( \underline{z}\right) =z_{n}~\sum_{\ell
=1;\left( n\right) }^{N}\left( \frac{1}{z_{n}-z_{\ell }}\right) +\sum_{\ell
=1;\left( n\right) }^{N}\left( \frac{z_{\ell }}{z_{n}-z_{\ell }}\right) 
\notag \\
&=&N-1+2~\sigma _{n}^{\left( 1,1\right) }\left( \underline{z}\right) ~,
\label{gn1}
\end{eqnarray}

\begin{eqnarray}
&&g_{n}^{\left( 2\right) }\left( \underline{z}\right) =-\sum_{\ell =1;\left(
n\right) }^{N}\left( \frac{z_{n}+z_{\ell }}{z_{n}-z_{\ell }}\right)
-2~\sum_{\ell ,m=1;(n)}^{N}\left[ \frac{z_{m}^{2}~\left( z_{n}+z_{\ell
}\right) }{d_{n\left( \ell m\right) }^{\left( 3\right) }\left( \underline{z}%
\right) }\right]  \notag \\
&=&1-N+2~\left( N-3\right) ~\sigma _{n}^{\left( 1,1\right) }\left( 
\underline{z}\right) -3~\sigma _{n}^{\left( 2,2\right) }\left( \underline{z}%
\right) +3~\left[ \sigma _{n}^{\left( 1,1\right) }\left( \underline{z}%
\right) \right] ^{2}~,  \label{gn2}
\end{eqnarray}

\begin{eqnarray}
&&g_{n}^{\left( 3\right) }\left( \underline{z}\right) =\sum_{\ell =1;\left(
n\right) }^{N}\left( \frac{z_{n}+z_{\ell }}{z_{n}-z_{\ell }}\right)
+6~\sum_{\ell ,m=1;(n)}^{N}\left[ \frac{\left( z_{n}+z_{\ell }\right)
~z_{m}^{2}}{d_{n\left( \ell m\right) }^{\left( 3\right) }\left( \underline{z}%
\right) }\right]  \notag \\
&&+6~\sum_{\ell ,k,m=1;(n)}^{N}\left[ \frac{\left( z_{n}+z_{\ell }\right)
~z_{m}^{2}~z_{k}^{3}}{d_{n\left( \ell km\right) }^{\left( 4\right) }\left( 
\underline{z}\right) }\right]  \notag \\
&=&N-1-2~\left( 3~N-7\right) \sigma _{n}^{(1,1)}(\underline{z})-\frac{9}{4}%
~\left( N-7\right) ~\sigma _{n}^{(2,2)}(\underline{z})+6~\sigma _{n}^{(3,3)}(%
\underline{z})  \notag \\
&&+\frac{9}{4}~\left( N-7\right) ~~\left[ \sigma _{n}^{(1,1)}(\underline{z})%
\right] ^{2}-9~\sigma _{n}^{(1,1)}(\underline{z})~\sigma _{n}^{(2,2)}(%
\underline{z})+3~\left[ \sigma _{n}^{(1,1)}(\underline{z})\right] ^{3}~, 
\notag \\
&&
\end{eqnarray}%
and so on (via standard, but increasingly tedious, computations).

To prove the identities (\ref{Identities1}) we note first of all that (\ref%
{fn1}) coincides with (\ref{A.6a}) (via (\ref{ShortHandNotat})). Next---to
prove (\ref{Iden1}) with (\ref{Recfn})---we apply the operator $\left(
z~d/dz-N\right) $ to (the long-hand version of) (\ref{Iden1}), getting
thereby the following chain of equations: 
\end{subequations}
\begin{subequations}
\label{Sequence1}
\begin{eqnarray}
&&\left( z~\frac{d}{d~z}-N\right) ^{j+1}~\psi \left( z\right)  \notag \\
&=&\left( z~\frac{d}{d~z}-N\right) ~\left\{ \psi \left( z\right)
~\sum_{n=1}^{N}\left[ \left( z-z_{n}\right) ^{-1}~f_{n}^{\left( j\right)
}\left( \underline{z}\right) \right] \right\}
\end{eqnarray}%
\begin{eqnarray}
&=&\left[ \left( z~\frac{d}{d~z}-N\right) ~\psi \left( z\right) \right]
~\sum_{n=1}^{N}\left[ \left( z-z_{n}\right) ^{-1}~f_{n}^{\left( j\right)
}\left( \underline{z}\right) \right]  \notag \\
&&-\psi \left( z\right) ~\sum_{n=1}^{N}\left[ z~\left( z-z_{n}\right)
^{-2}~f_{n}^{\left( j\right) }\left( \underline{z}\right) \right]
\end{eqnarray}%
\begin{eqnarray}
&=&\psi \left( z\right) ~\left\{ \sum_{n,\ell =1}^{N}\left[ \left(
z-z_{n}\right) ^{-1}~\left( z-z_{\ell }\right) ^{-1}~z_{\ell }~f_{n}^{\left(
j\right) }\left( \underline{z}\right) \right] \right.  \notag \\
&&\left. -\sum_{n=1}^{N}\left[ z~\left( z-z_{n}\right) ^{-2}~f_{n}^{\left(
j\right) }\left( \underline{z}\right) \right] \right\}
\end{eqnarray}%
\begin{eqnarray}
&=&\psi \left( z\right) ~\left\{ \sum_{n,\ell =1;~\ell \neq n}^{N}\left[
\left( z-z_{n}\right) ^{-1}~\left( z-z_{\ell }\right) ^{-1}~z_{\ell
}~f_{n}^{\left( j\right) }\left( \underline{z}\right) \right] \right.  \notag
\\
&&\left. -\sum_{n=1}^{N}\left[ \left( z-z_{n}\right) ^{-1}~f_{n}^{\left(
j\right) }\left( \underline{z}\right) \right] \right\}
\end{eqnarray}%
\begin{eqnarray}
&=&\psi \left( z\right) ~\left\{ \sum_{n,\ell =1;~\ell \neq n}^{N}\left[
\left( z-z_{n}\right) ^{-1}~\left( z_{n}-z_{\ell }\right) ^{-1}~z_{\ell
}~f_{n}^{\left( j\right) }\left( \underline{z}\right) \right] \right.  \notag
\\
&&-\sum_{n,\ell =1;~\ell \neq n}^{N}\left[ \left( z-z_{\ell }\right)
^{-1}~\left( z_{n}-z_{\ell }\right) ^{-1}~z_{\ell }~f_{n}^{\left( j\right)
}\left( \underline{z}\right) \right]  \notag \\
&&\left. -\sum_{n=1}^{N}\left[ \left( z-z_{n}\right) ^{-1}~f_{n}^{\left(
j\right) }\left( \underline{z}\right) \right] \right\}
\end{eqnarray}%
\begin{eqnarray}
&=&\psi \left( z\right) ~\left\{ \sum_{n=1}^{N}\left[ \left( z-z_{n}\right)
^{-1}~\left\{ -f_{n}^{\left( j\right) }\left( \underline{z}\right) \right.
\right. \right.  \notag \\
&&\left. \left. \left. +\sum_{\ell =1;~\ell \neq n}^{N}\left( z_{n}-z_{\ell
}\right) ^{-1}~\left[ z_{\ell }~f_{n}^{\left( j\right) }\left( \underline{z}%
\right) +z_{n}~f_{\ell }^{\left( j\right) }\left( \underline{z}\right) %
\right] \right\} \right] \right\} ~.
\end{eqnarray}%
The first equality is implied by (the long-hand version of) (\ref{Iden1});
the second by standard differentiation; the third, by (the long-hand version
of) (\ref{A.6a}); the fourth obtains by taking advantage of the cancellation
of the terms featuring double poles; the fifth, by using the elementary
identity $\left( z-z_{n}\right) ^{-1}~\left( z-z_{\ell }\right) ^{-1}=\left[
\left( z-z_{n}\right) ^{-1}-\left( z-z_{\ell }\right) ^{-1}\right] ~\left(
z_{n}-z_{\ell }\right) ^{-1}$; and the sixth and last, by exchanging the two
dummy indices $n$ and $\ell $ in the second sum. The last equality is
clearly the long-hand version of (\ref{Iden1}) (with $j$ replaced by $j+1\,$
and with (\ref{Recfn})); the recursion (\ref{Recfn}) is thereby proven.

To prove (\ref{Iden2}) with (\ref{gn}) we apply the operator $d/dz$ to (the
long-hand version of) (\ref{Iden1}), getting thereby the following chain of
equations: 
\end{subequations}
\begin{subequations}
\label{Sequence2}
\begin{eqnarray}
&&\left( \frac{d}{d~z}\right) ~\left( z~\frac{d}{d~z}-N\right) ^{j}~\psi
\left( z\right)  \notag \\
&=&\left( \frac{d}{d~z}\right) ~\left\{ \psi \left( z\right) ~\sum_{n=1}^{N}%
\left[ \left( z-z_{n}\right) ^{-1}~f_{n}^{\left( j\right) }\left( \underline{%
z}\right) \right] \right\}
\end{eqnarray}%
\begin{eqnarray}
&=&\left[ \frac{d}{d~z}~\psi \left( z\right) \right] ~\sum_{n=1}^{N}\left[
\left( z-z_{n}\right) ^{-1}~f_{n}^{\left( j\right) }\left( \underline{z}%
\right) \right]  \notag \\
&&-\psi \left( z\right) ~\sum_{n=1}^{N}\left[ \left( z-z_{n}\right)
^{-2}~f_{n}^{\left( j\right) }\left( \underline{z}\right) \right]
\end{eqnarray}%
\begin{eqnarray}
&=&\psi \left( z\right) ~\sum_{n,\ell =1}^{N}\left\{ \left[ \left(
z-z_{n}\right) ^{-1}~\left( z-z_{\ell }\right) ^{-1}\right] ~f_{n}^{\left(
j\right) }\left( \underline{z}\right) \right\}  \notag \\
&&-\psi \left( z\right) ~\sum_{n=1}^{N}\left[ \left( z-z_{n}\right)
^{-2}~f_{n}^{\left( j\right) }\left( \underline{z}\right) \right]
\end{eqnarray}%
\begin{equation}
=\psi \left( z\right) ~\sum_{n,\ell =1;~\ell \neq n}^{N}\left[ \left(
z-z_{n}\right) ^{-1}~\left( z-z_{\ell }\right) ^{-1}~f_{n}^{\left( j\right)
}\left( \underline{z}\right) \right]
\end{equation}%
\begin{equation}
=\psi \left( z\right) ~\sum_{n,\ell =1;~\ell \neq n}^{N}\left\{ \left[
\left( z-z_{n}\right) ^{-1}-\left( z-z_{\ell }\right) ^{-1}\right] ~\left(
z_{n}-z_{\ell }\right) ^{-1}~f_{n}^{\left( j\right) }\left( \underline{z}%
\right) \right\}
\end{equation}%
\begin{equation}
=\psi \left( z\right) ~\sum_{n,\ell =1;~\ell \neq n}^{N}\left\{ \left(
z-z_{n}\right) ^{-1}~\left( z_{n}-z_{\ell }\right) ^{-1}~\left[
f_{n}^{\left( j\right) }\left( \underline{z}\right) +f_{\ell }^{\left(
j\right) }\left( \underline{z}\right) \right] \right\} ~.
\end{equation}%
This sequence, (\ref{Sequence2}), of equalities is sufficiently analogous to
that reported above, see (\ref{Sequence1}), not to require any repetition of
the explanations provided above (after (\ref{Sequence1})). And clearly the
last equality proves the identity (\ref{Iden2}) with (\ref{gn}).

Finally, let us report explicit expressions of the quantities $%
f_{n,m}^{\left( j\right) }\left( \underline{\zeta }\right) $ and $%
g_{n,m}^{\left( j\right) }\left( \underline{\zeta }\right) ,$ see (\ref%
{fgjnm}) (conveniently obtained via (\ref{sigmanm}) from the expressions of $%
f_{n}^{\left( j\right) }\left( \underline{\zeta }\right) $ and $%
g_{n}^{\left( j\right) }\left( \underline{\zeta }\right) $ in terms of the
quantities $\sigma _{n}^{\left( r,\rho \right) }\left( \underline{\zeta }%
\right) $, see (\ref{ff}) and (\ref{ggg})): 
\end{subequations}
\begin{subequations}
\label{fnm}
\begin{equation}
f_{n,m}^{\left( 1\right) }\left( \underline{\zeta }\right) =\delta _{nm}~,
\end{equation}%
\begin{eqnarray}
f_{n,m}^{\left( 2\right) }\left( \underline{\zeta }\right) &=&-\delta _{nm}~%
\left[ 1+2~\sigma _{n}^{\left( 2,2\right) }\left( \underline{\zeta }\right) %
\right]  \notag \\
&&+\left( 1-\delta _{nm}\right) ~2~\left( \frac{\zeta _{n}}{\zeta _{n}-\zeta
_{m}}\right) ^{2}~,
\end{eqnarray}%
\begin{eqnarray}
&&f_{n,m}^{\left( 3\right) }\left( \underline{\zeta }\right) =\delta
_{nm}~\left\{ 1+9~\sigma _{n}^{\left( 2,2\right) }\left( \underline{\zeta }%
\right) +6~\sigma _{n}^{\left( 3,3\right) }\left( \underline{\zeta }\right)
\right.  \notag \\
&&\left. -3~\sigma _{n}^{\left( 1,1\right) }\left( \underline{\zeta }\right)
~\left[ \sigma _{n}^{\left( 1,1\right) }\left( \underline{\zeta }\right)
+2~\sigma _{n}^{\left( 2,2\right) }\left( \underline{\zeta }\right) \right]
\right\}  \notag \\
&&-6~\left( 1-\delta _{nm}\right) ~\frac{\zeta _{n}^{2}~\left[ \zeta
_{n}-\left( \zeta _{n}-\zeta _{m}\right) ~\sigma _{n}^{\left( 1,1\right)
}\left( \underline{\zeta }\right) \right] }{\left( \zeta _{n}-\zeta
_{m}\right) ^{3}}~,
\end{eqnarray}%
and so on (via standard, but increasingly tedious, computations);
respectively (see (\ref{gnj}) and (\ref{Identities2})) 
\end{subequations}
\begin{subequations}
\label{gnm}
\begin{equation}
g_{n,m}^{\left( 1\right) }\left( \underline{\zeta }\right) =-2~\delta
_{nm}~\sigma _{n}^{\left( 1,2\right) }\left( \underline{\zeta }\right)
+\left( 1-\delta _{nm}\right) ~\frac{2~\zeta _{n}}{\left( \zeta _{n}-\zeta
_{m}\right) ^{2}}~,
\end{equation}

\begin{eqnarray}
&&g_{n,m}^{\left( 2\right) }\left( \underline{\zeta }\right) =\delta
_{nm}~\left\{ -2~\left( N-3\right) ~\sigma _{n}^{\left( 1,2\right) }\left( 
\underline{\zeta }\right) +6~\sigma _{n}^{\left( 2,3\right) }\left( 
\underline{\zeta }\right) \right.  \notag \\
&&\left. -6~\sigma _{n}^{\left( 1,1\right) }\left( \underline{\zeta }\right)
~\sigma _{n}^{\left( 1,2\right) }\left( \underline{\zeta }\right) \right\} 
\notag \\
&&+\left( 1-\delta _{nm}\right) ~\left\{ \frac{2~\zeta _{n}~\left[ \left(
N-3\right) ~\zeta _{n}-N~\zeta _{m}\right] }{\left( \zeta _{n}-\zeta
_{m}\right) ^{3}}+\frac{6~\zeta _{n}~\sigma _{n}^{\left( 1,1\right) }\left( 
\underline{\zeta }\right) }{\left( \zeta _{n}-\zeta _{m}\right) ^{2}}%
\right\} ~,
\end{eqnarray}

\begin{eqnarray}
&&g_{n,m}^{\left( 3\right) }\left( \underline{\zeta }\right) =\delta
_{nm}~\left\{ 2~\left( 3~N-7\right) ~\sigma _{n}^{\left( 1,2\right) }\left( 
\underline{\zeta }\right) \right.  \notag \\
&&+\frac{9}{2}~\left( N-7\right) ~\sigma _{n}^{\left( 2,3\right) }\left( 
\underline{\zeta }\right) -18~\sigma _{n}^{\left( 3,4\right) }\left( 
\underline{\zeta }\right)  \notag \\
&&-\frac{9}{2}~\left( N-7\right) ~\sigma _{n}^{\left( 1,1\right) }\left( 
\underline{\zeta }\right) ~\sigma _{n}^{\left( 1,2\right) }\left( \underline{%
\zeta }\right) +9~\sigma _{n}^{\left( 1,2\right) }\left( \underline{\zeta }%
\right) ~\sigma _{n}^{\left( 2,2\right) }\left( \underline{\zeta }\right) 
\notag \\
&&\left. +18~\sigma _{n}^{\left( 1,1\right) }\left( \underline{\zeta }%
\right) ~\sigma _{n}^{\left( 2,3\right) }\left( \underline{\zeta }\right) -9~%
\left[ \sigma _{n}^{\left( 1,1\right) }\left( \underline{\zeta }\right) %
\right] ^{2}~\sigma _{n}^{\left( 1,2\right) }\left( \underline{\zeta }%
\right) \right\}  \notag \\
&&+\left( 1-\delta _{nm}\right) ~\zeta _{n}~\left\{ \frac{18~\zeta _{m}^{2}~%
}{\left( \zeta _{n}-\zeta _{m}\right) ^{4}}-\frac{9~\zeta _{m}~\left[ \left(
N-7\right) /2~+2~\sigma _{n}^{\left( 1,1\right) }\left( \underline{\zeta }%
\right) \right] }{\left( \zeta _{n}-\zeta _{m}\right) ^{3}}\right.  \notag \\
&&\left. \frac{-2~\left( 3~N-7\right) +\left( 9/2\right) ~\left( N-7\right)
~\sigma _{n}^{\left( 1,1\right) }\left( \underline{\zeta }\right) -9~\sigma
_{n}^{\left( 2,2\right) }\left( \underline{\zeta }\right) +9~\left[ \sigma
_{n}^{\left( 1,1\right) }\left( \underline{\zeta }\right) \right] ^{2}}{%
\left( \zeta _{n}-\zeta _{m}\right) ^{2}}\right\} ~,  \notag \\
&&
\end{eqnarray}%
and so on (via standard, but increasingly tedious, computations).

\bigskip

\end{subequations}

\end{document}